\documentclass[11pt,a4paper]{article}
\usepackage{jcappub}
\usepackage{lmodern}
\usepackage{graphicx,color}
\graphicspath{{figures/}}
\usepackage{comment}
\usepackage{cancel}

\usepackage[T1]{fontenc}
\usepackage[utf8]{inputenc}
\usepackage{color}
\usepackage{amsmath}
\usepackage{amssymb}
\usepackage{esint}
\usepackage{appendix}
\usepackage{bm}
\usepackage{url}
\usepackage{hyperref}
\usepackage{xcolor}
\usepackage[normalem]{ulem}

\makeatletter

\definecolor{somegreen}{cmyk}{0,0.49,0.98,0.09}
\definecolor{red}{rgb}{1,0,0}
\definecolor{magenta}{cmyk}{0,1,0,0}
\definecolor{lavender}{cmyk}{0.16,0.67,0,0.57}
\definecolor{darkgreen}{rgb}{0,0.65,0.05}
\definecolor{antiquefuchsia}{rgb}{0.33, 0.1, 0.89}

\def\beqra{\begin{eqnarray}}
\def\eeqra{\end{eqnarray}}
\def\beq{\begin{equation}}
\def\eeq{\end{equation}}

\def\vp{\bar{\varphi}}

\def\L{\Lambda}

\def\vp{\varphi}

\def\bx{{\bf{x}}}
\def\bk{{\bf{k}}}
\def\bp{{\bf{p}}}
\def\bq{{\bf{q}}}

\def\bn{{\bf{n}}}

\def\bV0{{\bf{V_0}}}

\def\agt{~\mbox{\raisebox{-.6ex}{$\stackrel{>}{\sim}$}}~}

\def\bx{{\bf{x}}}

\def\bk{{\bf{k}}}
\def\bp{{\bf{p}}}
\def\bq{{\bf{q}}}

\def\bV{{\bf{V}}}

\def\vp{\varphi}

\newcommand{\hMpc}{h\,\mathrm{Mpc}^{-1}}

\newcommand{\MP}[1]{{\textcolor{red}{\bf #1}}\marginpar{\textcolor{red}{$\bullet$ Matteo}}}

\let\jnl@style=\rm
\def\ref@jnl#1{{\jnl@style#1}}

\def\aj{\ref@jnl{AJ}}                   % Astronomical Journal
\def\actaa{\ref@jnl{Acta Astron.}}      % Acta Astronomica
\def\araa{\ref@jnl{ARA\&A}}             % Annual Review of Astron and Astrophys
\def\apj{\ref@jnl{ApJ}}                 % Astrophysical Journal
\def\apjl{\ref@jnl{ApJ}}                % Astrophysical Journal, Letters
\def\apjs{\ref@jnl{ApJS}}               % Astrophysical Journal, Supplement
\def\ao{\ref@jnl{Appl.~Opt.}}           % Applied Optics
\def\apss{\ref@jnl{Ap\&SS}}             % Astrophysics and Space Science
\def\aap{\ref@jnl{A\&A}}                % Astronomy and Astrophysics
\def\aapr{\ref@jnl{A\&A~Rev.}}          % Astronomy and Astrophysics Reviews
\def\aaps{\ref@jnl{A\&AS}}              % Astronomy and Astrophysics, Supplement
\def\azh{\ref@jnl{AZh}}                 % Astronomicheskii Zhurnal
\def\baas{\ref@jnl{BAAS}}               % Bulletin of the AAS
\def\bac{\ref@jnl{Bull. astr. Inst. Czechosl.}}
\def\caa{\ref@jnl{Chinese Astron. Astrophys.}}
\def\cjaa{\ref@jnl{Chinese J. Astron. Astrophys.}}
\def\icarus{\ref@jnl{Icarus}}           % Icarus
\def\jcap{\ref@jnl{J. Cosmology Astropart. Phys.}}
\def\jrasc{\ref@jnl{JRASC}}             % Journal of the RAS of Canada
\def\memras{\ref@jnl{MmRAS}}            % Memoirs of the RAS
\def\mnras{\ref@jnl{MNRAS}}             % Monthly Notices of the RAS
\def\na{\ref@jnl{New A}}                % New Astronomy
\def\nar{\ref@jnl{New A Rev.}}          % New Astronomy Review
\def\pra{\ref@jnl{Phys.~Rev.~A}}        % Physical Review A: General Physics
\def\prb{\ref@jnl{Phys.~Rev.~B}}        % Physical Review B: Solid State
\def\prc{\ref@jnl{Phys.~Rev.~C}}        % Physical Review C
\def\prd{\ref@jnl{Phys.~Rev.~D}}        % Physical Review D
\def\pre{\ref@jnl{Phys.~Rev.~E}}        % Physical Review E
\def\prl{\ref@jnl{Phys.~Rev.~Lett.}}    % Physical Review Letters
\def\pasa{\ref@jnl{PASA}}               % Publications of the Astron. Soc. of Australia
\def\pasp{\ref@jnl{PASP}}               % Publications of the ASP
\def\pasj{\ref@jnl{PASJ}}               % Publications of the ASJ
\def\rmxaa{\ref@jnl{Rev. Mexicana Astron. Astrofis.}}%
\def\qjras{\ref@jnl{QJRAS}}             % Quarterly Journal of the RAS
\def\skytel{\ref@jnl{S\&T}}             % Sky and Telescope
\def\solphys{\ref@jnl{Sol.~Phys.}}      % Solar Physics
\def\sovast{\ref@jnl{Soviet~Ast.}}      % Soviet Astronomy
\def\ssr{\ref@jnl{Space~Sci.~Rev.}}     % Space Science Reviews
\def\zap{\ref@jnl{ZAp}}                 % Zeitschrift fuer Astrophysik
\def\nat{\ref@jnl{Nature}}              % Nature
\def\iaucirc{\ref@jnl{IAU~Circ.}}       % IAU Cirulars
\def\aplett{\ref@jnl{Astrophys.~Lett.}} % Astrophysics Letters
\def\apspr{\ref@jnl{Astrophys.~Space~Phys.~Res.}}
\def\bain{\ref@jnl{Bull.~Astron.~Inst.~Netherlands}}
\def\fcp{\ref@jnl{Fund.~Cosmic~Phys.}}  % Fundamental Cosmic Physics
\def\gca{\ref@jnl{Geochim.~Cosmochim.~Acta}}   % Geochimica Cosmochimica Acta
\def\grl{\ref@jnl{Geophys.~Res.~Lett.}} % Geophysics Research Letters
\def\jcp{\ref@jnl{J.~Chem.~Phys.}}      % Journal of Chemical Physics
\def\jgr{\ref@jnl{J.~Geophys.~Res.}}    % Journal of Geophysics Research
\def\jqsrt{\ref@jnl{J.~Quant.~Spec.~Radiat.~Transf.}}
\def\memsai{\ref@jnl{Mem.~Soc.~Astron.~Italiana}}
\def\nphysa{\ref@jnl{Nucl.~Phys.~A}}   % Nuclear Physics A
\def\physrep{\ref@jnl{Phys.~Rep.}}   % Physics Reports
\def\physscr{\ref@jnl{Phys.~Scr}}   % Physica Scripta
\def\planss{\ref@jnl{Planet.~Space~Sci.}}   % Planetary Space Science
\def\procspie{\ref@jnl{Proc.~SPIE}}   % Proceedings of the SPIE

\makeatother

\title{Renormalized Perturbation Theory at Field-level: the LSS bootstrap in \texttt{GridSPT}}

\author[a]{Matteo Peron,}
\author[b,c,d]{Takahiro Nishimichi,}
\author[a]{Massimo Pietroni,}
\author[c,d]{and Atsushi Taruya}
\affiliation[a]{Dipartimento di Scienze Matematiche, Fisiche e Informatiche, Universit\`a di Parma, and INFN Gruppo Collegato di Parma, Parco Area delle Scienze 7/A, I-43124, Parma, Italy}
\affiliation[b]{Department of Astrophysics and Atmospheric Sciences, Faculty of Science, Kyoto Sangyo University, Motoyama, Kamigamo, Kita-ku, Kyoto,Kyoto 603-8555, Japan}
\affiliation[c]{Center for Gravitational Physics and Quantum Information,Yukawa Institute for Theoretical Physics, Kyoto University, Kyoto 606-8502, Japan}
\affiliation[d]{Kavli Institute for the Physics and Mathematics of the Universe (WPI),UTIAS, The University of Tokyo, 5-1-5 Kashiwanoha, Kashiwa, Chiba 277-8583, Japan}

\abstract{
We present a first step toward field-level cosmological inference beyond the standard $\Lambda$CDM model, focusing on optimizing precision tests in the nonlinear regime of large-scale structure (LSS). As an illustrative case, we study the model-independent ``bootstrap'' coefficient of the second-order perturbation theory (PT) kernel for matter in real space, which we use as a proxy for new physics effects in the nonlinear sector. We discuss in details the ultraviolet (UV) cutoff dependence induced by discretizing fields on a grid, which requires proper renormalization to eliminate grid artifacts. We formulate a Wilsonian perturbative framework in which the evolution from a UV theory defined at a high cutoff $\Lambda_{\rm uv}$ down to lower cutoffs is computed analytically, even beyond the validity of a derivative expansion. Within this framework, we develop an extended version of the \texttt{GridSPT} code incorporating the bootstrap parameterization and demonstrate how cutoff-independent predictions can be achieved through the inclusion of appropriate counterterms. We validate our approach at third- and fifth-order in PT, emphasizing the importance of higher-derivative contributions for unbiased parameter extraction. Our framework is readily extendable to biased tracers and redshift-space distortions.
}

\emailAdd{matteo.peron@unipr.it} 
\emailAdd{massimo.pietroni@unipr.it}
\emailAdd{ataruya@yukawa.kyoto-u.ac.jp}
\emailAdd{nishimichi@cc.kyoto-su.ac.jp}
\begin{document}

\begin{flushright}
YITP-25-88
\end{flushright}
\maketitle

\section{Introduction}
\label{sect:intro}
The large-scale structure (LSS) of the Universe  offers a critical testing ground for our understanding of fundamental physics. In order to fully exploit data from surveys like \textsc{DESI} \cite{DESI:2016fyo} and
\textsc{Euclid}  \cite{Amendola:2016saw, Euclid:2024yrr}
it becomes increasingly important to develop methods that optimize the extraction of cosmological information from LSS, particularly in ways that are sensitive to physics beyond the standard $\L$CDM model. Various ``new physics'' scenarios have been suggested, but currently there is no theoretical basis to select one over the others. Additionally, in most plausible cases, the predicted signatures are quite elusive, underscoring the need for a model-independent framework that can identify minor anomalies. To this end, a ``LSS bootstrap'' approach was formulated in \cite{DAmico:2021rdb}, and further developed in \cite{Marinucci:2024add, Ansari:2025nsf}. It is a symmetry-based approach in which no specific form for the Lagrangian and the equations of motion is assumed. The most general form for the dark matter (DM) density and velocity fields, as well as for biased tracers (galaxies, DM halos, ...) compatible with the assumed symmetries, is derived at each order in perturbation theory (PT) in terms of the linear field and a finite number of free time-dependent coefficients. The detection of deviations of these coefficients from their $\L$CDM values would then be a signal of new physics.  

While state of the art analyses frequently concentrate on low-order correlators like the power spectrum \cite{DAmico:2019fhj, Ivanov:2019pdj} and bispectrum \cite{DAmico:2022osl,Ivanov:2023qzb}, recent investigations have explored novel summary statistics as a means to extract non-Gaussian information encapsulated in higher-order correlators \cite{MoradinezhadDizgah:2019xun,Philcox:2020fqx, Massara:2020pli, Coulton:2023ouk, Valogiannis:2021chp, Eickenberg:2022qvy, Valogiannis:2023mxf, Peron:2024xaw, Marinucci:2024bdq} (for a review, see \cite{Beyond-2pt:2024mqz}). A natural progression in this sequence includes conducting inference directly at the field level, thus preserving all accessible informational content in an uncompressed form.
This approach, which has been developed in \cite{Jasche2013,Kitaura2013,Wang2013,Wang2014,Jasche2019,Lavaux2019,Schmidt2019,Modi2022,Kostic2023,Andrews2023,Bayer2023,Doeser2024,Nguyen2024}
 and further investigated in \cite{ McQuinn2021,Leclercq2021,Porqueres2022,Porth2023,Cabass2024,Schmidt:2025iwa},
 has the potential of providing an ideal setting for the exploration of both linear and non-linear manifestations of new physics. 

In this work, we pursue the construction of a field-level perturbative framework for model-independent cosmological inference, emphasizing its application to precision constraints on departures from  $\L$CDM.

To this end, we build upon the \texttt{GridSPT} code \cite{Taruya2018,Taruya2021,Taruya2022}, an implementation of Eulerian PT (for a review, see \cite{PT}) on discretized grids, and modify it to include the LSS bootstrap parametrization, in order to systematically explore deviations from the standard model at the nonlinear level. 

Working at the field level on a grid necessarily introduces an ultraviolet (UV) cutoff. The linear fields which form the building blocks of higher PT orders contain wavenumbers only up to the Nyquist frequency of the grid, or up to even lower values, as we will see when discussing the need to avoid spurious effects such as \textit{aliasing}. A meticulous treatment of this artificial cutoff-dependence is therefore essential to obtain physically meaningful, cutoff-independent, results.

We take a Wilsonian perspective (for a similar philosophy, see \cite{Carroll:2013oxa} and, more recently, \cite{Rubira:2023vzw, Nikolis:2024kbx}, where also biased tracers and non-Gaussian initial conditions were considered): the starting point is the  definition of the perturbative model at a high momentum cutoff scale, $\L_{\rm uv}$. The model is built from the linear field coarse-grained at the scale $\L_{\rm uv}$, and is given by a truncated PT series plus a set of ``counterterms''. As in the Effective Field Theory of LSS (EFTofLSS) \cite{Baumann:2010tm, Pietroni:2011iz, Carrasco:2012cv}, the form of the counterterms is determined by the symmetries of the system (rotational invariance, Extended Galilean Invariance (EGI), and, for the matter and velocity fields, mass and momentum conservation). Their number, on the other hand, is determined by the order of the approximation, both in the PT expansion and in the expansion on derivatives of the fields. The coefficients of the counterterms encode physics beyond the cutoff, that is, from scales $q> \L_{\rm uv}$ not described by the model, and therefore are not computable. Their values must be fixed by  {\it renormalization conditions}, provided by the comparison between the computed observables and the measured ones.  On the other hand, the {\it running} of these parameters, that is, their evolution as the cutoff scale is lowered to $\L<\L_{\rm uv}$, is computable in PT. Therefore, the requirement that physical observables are cutoff-independent provides the connection between the UV model and any other model defined at a lower scale $\L$.  We perform this renormalization explicitly up to fifth order in perturbation theory, achieving properly renormalized field-level models suitable for the determination of the bootstrap parameters.  This procedure allows us not only to maintain theoretical consistency but also to work efficiently with smaller grids in practical computations.

In particular, we highlight the crucial role of higher-derivative terms in achieving precision inference. These terms, which emerge naturally in the renormalization procedure, can be computed analytically within our framework. Including them is essential to accurately capture small-scale physics and to ensure the robust determination of bootstrap parameters necessary for model-independent tests of cosmology.

The paper is organized as follows. In Sect.~\ref{sect:bootstrap} we give the explicit forms for the bootstrap PT kernel at second order, both in Fourier and in configuration space. In Sect.~\ref{sect:GridSPT} we review the \texttt{GridSPT} code and describe the modifications implemented in order to compute the bootstrap kernels. In Sect.~\ref{sect:LambdaDep} we discuss the need of a momentum cutoff when formulating field-theory on a grid, and show in detail how different models, defined at different values of the cutoff, are related (running). In Sect.~\ref{sect:Likelihood} we compare our PT models with N-body simulations and show how to extract the bootstrap parameters. In Sect.~\ref{sect:results} we present the results of the analysis and verify the proper renormalization of the models, and finally, in Sect.~\ref{Sect:discussion} we discuss our results and give our conclusions. Moreover, in Appendix \ref{app:third} we discuss the effect of the bootstrap parameter when included also in the third order PT contribution, and in Appendix \ref{app:noise} we give some details on the implementation of artificial noise in our dark matter simulations.

\section{Field level bootstrap}
\label{sect:bootstrap}
In this section we review the perturbative model that we will implement in this analysis, based on the ``LSS bootstrap'' approach; for more details, see \cite{DAmico:2021rdb, Piga:2022mge, Marinucci:2024add}.
\subsection{First order}
We will express perturbations in terms of the density contrast, $\delta(\bx,a)$, and of the rescaled velocity divergence, 
\beq
\theta(\bx,a)\equiv \partial_i u^i(\bx,a)\,,
%\quad u^i\equiv -\frac{v^i}{f a H}\,,
\eeq
where 
\beq
u^i(\bx,a)\equiv -\frac{ v^i(\bx,a)}{f a H(a)}\,,
\eeq
$v^i(\bx, a)$ is the peculiar velocity, 
$f=d \log D_+/d \log a$, is the growth function, and $D_+(a)$ the linear growth factor, that we will assume to be scale-independent.
We then expand the fields according to the standard perturbative Ansatz \cite{Bernardeau_2002},
\beq
\label{PTexp}
\delta^{[N], {\rm PT}}(\bx,a)=\sum_{n=1}^N \delta^{(n)}(\bx,a)\,,\quad \theta^{[N], {\rm PT}}(\bx,a)=\sum_{n=1}^N \theta^{(n)}(\bx,a)\,.
%\quad {\bf u}(\bx,\eta)=\sum e^{n \eta} \,{\bf u}_n(\bx,\eta)\,,\quad 
\eeq
In the following, we will indicate with lowercase letters inside parentheses (e.g. like in $\delta^{(n)}$) the perturbative contribution at a given order, and with capital letters inside square brackets (e.g. like in $\delta^{[N], {\rm PT}}$), the order of truncation of the PT series, that is, the sum of all the $m$-th order contributions with $1\le m\le N$.

The first order solutions on the growing mode are
\beq
\label{eq:PT1st}
\delta^{(1)}(\bx,a)=\theta^{(1)}(\bx,a)\equiv \varphi(\bx,a)\,,
\eeq
with $\varphi(\bx,a)=D_+(a) \vp_0(\bx)$ the linear density field, a Gaussian realization whose variance is given by the linear power spectrum, $P(k)$, of the chosen cosmology.
\subsection{Second order}
In Fourier space, the second order fields can be expressed in terms of convolutions, 
\begin{align}
\delta^{(2)}(\bk,a) &= {\cal I}_{\bk;\bq_1,\bq_2} F_{2}(\bq_1,\bq_2; a) \varphi(\bq_1,a)\varphi(\bq_2,a)\,,\label{eq:delta2}\\
\theta^{(2)}(\bk,a) &= {\cal I}_{\bk;\bq_1,\bq_2} G_{2}(\bq_1,\bq_2; a) \varphi(\bq_1,a)\varphi(\bq_2,a)\,,
\label{eq:theta2}
\end{align}
where
\beq
{\cal I}_{\bk;\bq_1,\cdots,\bq_n}\equiv\int\frac{d^3 q_1\cdots d^3q_n}{(2\pi)^{3(n-1)}}\delta_D(\bk-\bq_1\cdots-\bq_n)\,,
\eeq
and the convolution kernels are,
\begin{align}
F_{2}(\bq_1,\bq_2; a)  & =  \beta(\bq_{1},\bq_{2})+ \frac{a_{\gamma}^{(2)}(a)}{2}\,\gamma(\bq_{1},\bq_{2})\,,\label{eq:F2}\\
G_{2}(\bq_1,\bq_2;a)  & =  \beta(\bq_{1},\bq_{2})+ \frac{d_{\gamma}^{(2)}(a)}{2}\,\gamma(\bq_{1},\bq_{2})\,,\label{eq:G2}
\end{align}
with the mode-coupling functions  defined as
\begin{align}
\beta(\bq_{1},\bq_{2}) & =\frac{|\bq_{1}+\bq_{2}|^{2}\bq_{1}\cdot \bq_{2}}{2q_{1}^{2}q_{2}^{2}}\,,\\
\gamma(\bq_{1},\bq_{2}) & =1-\frac{(\bq_{1}\cdot \bq_{2})^{2}}{q_{1}^{2}q_{2}^{2}}\,.
\end{align}
The coefficients of the $\beta(\bq_1,\bq_2)$ couplings in Eqs.~\eqref{eq:F2} and \eqref{eq:G2}, are fixed to unity by EGI. On the other hand, the bootstrap coefficients $a^{(2)}_\gamma$ and $d^{(2)}_\gamma$ are not constrained by symmetries, but assume different, in general time-dependent, values in different models.  
As an example, in $\Lambda$CDM the equations for the second order bootstrap coefficients are \cite{DAmico:2021rdb}
\begin{align}
\label{adev}
&\frac{d\, a_{\gamma}^{(2)}(a) }{d\ln{a}} = f \left(2 - 2 a_{\gamma}^{(2)}(a) + d_{\gamma}^{(2)}(a) \right)\,, \\
& \frac{d\, d_{\gamma}^{(2)}(a)}{d\ln{a}} = f\left(\frac{3}{2}\frac{\Omega_m(a)}{f^2} \bigg(a_{\gamma}^{(2)}(a) - d_{\gamma}^{(2)}(a) \bigg) - d_{\gamma}^{(2)}(a)\right)\,.\nonumber
\end{align}
Deviations from the solutions of these equations parameterize deviations from $\L$CDM in a model-independent way. 
The standard approach in PT is to use the ``Einstein-de Sitter (EdS) limit'' for the kernels,\footnote{For analyses going beyond the EdS approximation, see \cite{Pietroni:2008jx, Donath:2020abv, Piga:2022mge}.} obtained by setting  $\Omega_m/f^2=1$, in the above equations, which then give the time-independent values \cite{Bernardeau_2002}
\beq
\quad a_{\gamma,\;{\rm EdS}}^{(2)} = \frac{10}{7}\,,\quad 
d_{\gamma,\;{\rm EdS}}^{(2)} = \frac{6}{7}\,.
\label{eq:EdSvals}
\eeq
These values provide the initial conditions for the equations in all models that reduce to EdS at early times.

The bootstrap approach can be extended beyond second order, with an increasing number of cosmology-dependent coefficients. For the explicit expressions up to third order in the Eulerian formalism, see Appendix \ref{app:third} and \cite{DAmico:2021rdb, Piga:2022mge}, and for those up to sixth order in the equivalent Lagrangian formalism see \cite{Marinucci:2024add}. Moreover, the program can be extended from the matter and velocity fields to the density contrast for biased tracers. In this case, mass and momentum conservation are not imposed, therefore the kernels contain an higher number of free parameters, corresponding to those of the perturbative bias expansion (see, for instance, \cite{Desjacques:2016bnm}).  In this paper, we will restrict the analysis to the matter density field in real space. Moreover, we will set all coefficients of the kernels of order higher than second  to their EdS values, and will leave only the second order bootstrap coefficient, $a_{\gamma}^{(2)}$, free to vary.

In configuration space the second order density field is given by,
\begin{align}
   \delta^{(2)}(\bx)&= \varphi^{(2)}_\beta(\bx)+\frac{a_{\gamma}^{(2)}}{2}\varphi^{(2)}_\gamma(\bx) \,,\label{delta2} 
\end{align}
where we have omitted the time dependence and have defined the second order fields,
\begin{align}
\varphi^{(2)}_\beta(\bx)&\equiv\partial_i \varphi(\bx) \left(\frac{\partial_i}{\partial^2}\varphi(\bx)\right)+\left(\frac{\partial_i\partial_j}{\partial^2}\varphi(\bx)\right)\left(\frac{\partial_i\partial_j}{\partial^2}\varphi(\bx)\right)\,,
\label{eq:phibeta}\\
\varphi^{(2)}_\gamma(\bx)&\equiv\left(\varphi(\bx)\right)^2- \left(\frac{\partial_i\partial_j}{\partial^2}\varphi(\bx)\right)\left(\frac{\partial_i\partial_j}{\partial^2}\varphi(\bx)\right)\,.
\label{eq:phigamma}
\end{align}

\section{\texttt{GridSPT} and bootstrap}
\label{sect:GridSPT}
\subsection{The \texttt{GridSPT} code} Our numerical implementation of the bootstrap perturbative approach is based on the grid-based Eulerian PT code called \texttt{GridSPT} \citep{Taruya2018,Taruya2021,Taruya2022}. Specifically, the code performs standard PT calculations at the field level, and generates numerical realizations of higher-order density and velocity fields at each grid point. Notably, by applying a novel $n$EPT ($n$th-order Eulerian PT) scheme, the field-level calculation with \texttt{GridSPT} also yields improved predictions for summary statistics, outperforming those from the conventional order-by-order PT calculations \cite{Wang2023,Wang2024}. 

%\MaxP{[please, add references on GridSPT, and expand/improve the description]}.
The core of the \texttt{GridSPT} algorithm is a real-space recursion relation under the 
EdS approximation. For $n\geq2$, 
the $n$-th order Eulerian density and velocity fields are constructed  in configuration space from their lower-order counterparts as follows \cite{Taruya2022}:

\begin{comment}
Under the assumption that dark matter (DM) behaves like an incompressible, self-gravitating fluid, its dynamics are described by the Euler and continuity equations coupled with the Poisson equation. In addition, under the assumption of irrotational flow, which is found to be valid at large scales \MP{cit}, the fully non-linear equations of motion further simplify to the form:

\begin{equation}
    \frac{d}{d\eta}\begin{pmatrix}
    \delta \\
    \theta
    \end{pmatrix}+\Omega_{ab}\begin{pmatrix}
    \delta \\
    \theta
    \end{pmatrix}=\begin{pmatrix}
    \partial_i\left(\delta u^i\right) \\
    \partial_i\left[\left(u_j\nabla^j\right)u^i\right]
    \end{pmatrix},
\end{equation}

where the matrix $\Omega_{ab}\equiv\Omega_{ab}(\eta)$ depends on cosmology (and time). If we condition the dynamics to an EdS Universe and perform a perturbative expansion of both the overdensity $\delta$ and divergence of the velocity field $\theta$, an iterative solution can be found in the form of:

\begin{equation}
\label{eq:gridspt_main}
    \begin{pmatrix}
    \delta^{(n)} \\
    \theta^{(n)}
    \end{pmatrix} = \frac{2}{(2n+3)(n-1)}\begin{pmatrix}
    n + \frac{1}{2} & 1 \\
    \frac{3}{2} & n
    \end{pmatrix}\sum_{m=1}^{n-1}
    \begin{pmatrix}
    \partial_i\delta^{(m)}\left[u^{(n-m)}\right]^i + \delta^{(m)} \theta^{(n-m)} \\
    \frac{1}{2} \partial^2 \left(\left[u^{(m)}\right]_i\left[u^{(n-m)}\right]^i\right)
    \end{pmatrix}.
\end{equation}
\end{comment}
%\MaxP{[Please check the equation below]}

\begin{equation}
    \begin{pmatrix}
    \delta^{(n)} \\
    \\
    \theta^{(n)}
    \end{pmatrix} = \frac{2}{(2n+3)(n-1)}\begin{pmatrix}
    n + \frac{1}{2} & 1 \\
    & \\
    \frac{3}{2} & n
    \end{pmatrix}\sum_{m=1}^{n-1}
    \begin{pmatrix}
    \partial_i\delta^{(m)}\frac{\partial_i}{\partial^2}\theta^{(n-m)} + \delta^{(m)} \theta^{(n-m)} \\
   & \\
    \frac{1}{2} \partial^2 \left(\frac{\partial_i}{\partial^2}\theta^{(m)}\frac{\partial_i}{\partial^2}\theta^{(n-m)}\right)
    \end{pmatrix}.
    \label{eq:iterspt}
\end{equation}
Given the linear fields $\delta^{(1)}(\bx)$ and $\theta^{(1)}(\bx)$ on grids as in Eq.~\eqref{eq:PT1st}, the code computes the nonlinear source terms given on the right-hand side of Eq.~\eqref{eq:iterspt}, making use of the Fast Fourier Transform. 
The differential operations involved in the source terms are performed via expressions such as 
\begin{align}
\frac{\partial_i}{\partial^2}\varphi(\bx)&=-i \int \frac{d^3 k}{(2\pi)^3}e^{i\bk\cdot\bx}\,\frac{k_i}{k^2} \varphi(\bk)\,,\\
\frac{\partial_i\partial_j}{\partial^2}\varphi(\bx)&=-\int \frac{d^3 k}{(2\pi)^3}e^{i\bk\cdot\bx}\,\frac{k_ik_j}{k^2} \varphi(\bk)\,.
\end{align}

\subsection{Bootstrapping \texttt{GridSPT}}
The fields computed from \texttt{GridSPT} by solving Eq.~\eqref{eq:iterspt} rely on the EdS approximation for the PT kernels. In order to compute the $n=2$ bootstrap density field,  we write Eq.~\eqref{eq:delta2} as
\begin{equation}
\delta^{(2)}(\bx)= \delta^{(2)}_{\rm EdS}(\bx) + \frac{1}{2} \varepsilon_\gamma \,a^{(2)}_{\gamma,{\rm EdS}} \,\vp^{(2)}_{\gamma}(\bx)\,,
\label{eq:delta2boot}
\end{equation}
where the parameter
\begin{equation}
    \varepsilon_{\gamma}\equiv \frac{a_{\gamma}^{(2)}}{a_{\gamma,{
{\rm EdS}}}^{(2)}}-1\,,
\end{equation}
represents the relative deviation from the EdS limit. 
Notice that, at $z=1$ and for the $\Lambda$CDM model considered in our analysis, we have $ \varepsilon_{\gamma,\L{\rm CDM}}\simeq - 7.8\cdot 10^{-4}$.

The field  $ \delta^{(2)}_{\rm EdS}(\bx)$ will be taken from the output of  \texttt{GridSPT}, while the field  $\varphi_{\gamma}^{(2)}$, defined in Eq.~\eqref{eq:phigamma}, will be built from the same linear field $\vp(\bx)$  as $ \delta^{(2)}_{\rm EdS}(\bx)$.
In order to reduce the order of the tensor structures involved, we write  Eq.~\eqref{eq:phigamma}, as %\MP{Should we specify the irrotational velocity field approximation?}
\begin{equation}
\varphi_{\gamma}^{(2)}(\bx)=\left(\vp(\bx)\right)^2-\frac{1}{2}\partial^2\left[\left(\frac{\partial_i}{\partial^2}\vp(\bx)\right)\left(\frac{\partial_i}{\partial^2}\vp(\bx)\right)\right]+\left(\frac{\partial_i}{\partial^2}\vp(\bx)\right)\partial_i \vp(\bx)\,.
\end{equation}
Since in this analysis we will focus on the $a_\gamma^{(2)}$ parameter, we will set all the other bootstrap parameters to their EdS value, and therefore, we will compute all the fields at orders higher than $n=2$ with \texttt{GridSPT}, that is,
\beq
\delta^{(n)}(\bx)= \delta^{(n)}_{\rm EdS}(\bx)\,,\qquad (n >2)\,.
\eeq
Actually, this is not completely consistent, as the relations found in \cite{DAmico:2021rdb} imply that all higher order PT kernels depend also on the second order coefficient $a_\gamma^{(2)}$. However, we checked explicitly that this extra dependence is largely subdominant and that taking it into account in the third order field does not change our results in any appreciable way, see Appendix \ref{app:third}.

\section{Field on the grid: regularization and renormalization}
\label{sect:LambdaDep}

\begin{figure}
    \centering
    \includegraphics[width=\textwidth]{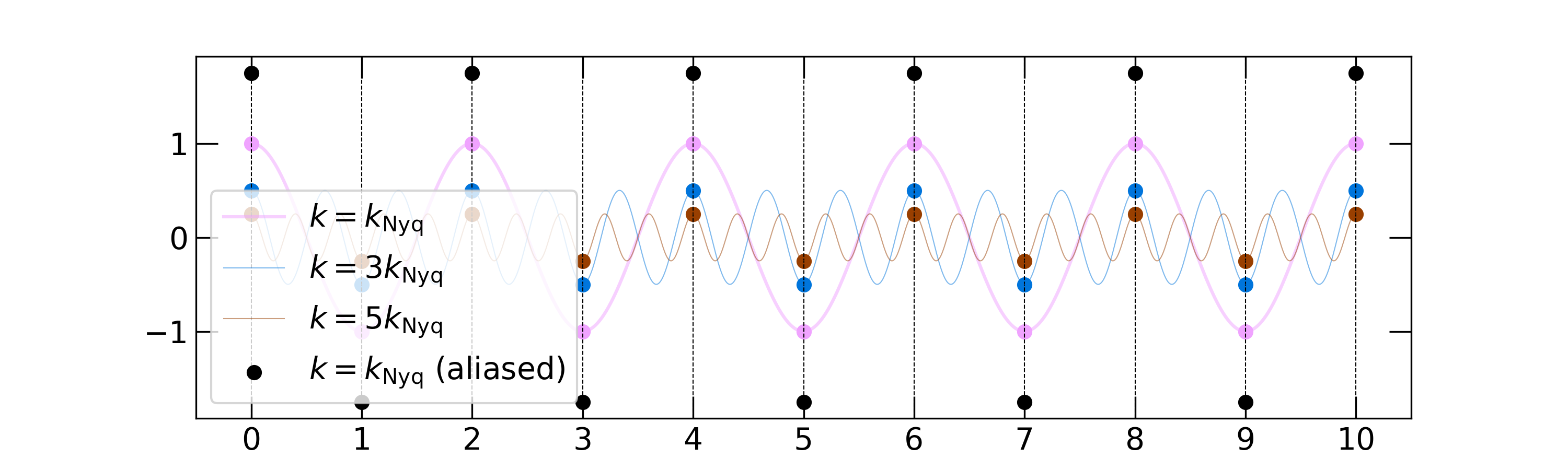}
    \caption{Simple illustration of how aliasing works. A detector samples a signal at each unit interval in a domain of length $L=10$, so $N_g=10$, meaning that the Nyquist frequency is $k_{\rm Nyq}=\frac{\pi N_g}{L}=\pi$. A wavemode at the Nyquist frequency (pink) is correctly sampled, however frequencies at $k=(2n+1)k_{\rm Nyq}$ (e.g., blue and brown) cannot be resolved and appear to the detector as $k=k_{\rm Nyq}$ as well. The result is that the measured signal at $k_{\rm Nyq}$ is the sum of all the contributions for $n>0$ (black).}
    \label{fig:aliasing}
\end{figure}
\subsection{The need for a momentum cutoff}
As discussed in \cite{Taruya:2021ftd}, when performing perturbative computations on a grid, one should take care in avoiding spurious ``aliasing'' effects on the resulting fields. These originate from products of  linear fields on the same grid point or, using a field-theoretical language, from {\it composite operators}.  On a finite cubic grid of  volume $L^3$ and  $N_g^3$ grid points a mode $\bk$ is indistinguishable from modes of the form $\bk + 2\, \bn\, k_{\rm Nyq}$, where $ k_{\rm Nyq}= \pi N_g/L$ is the  Nyquist frequency and $\bn=(n_x,n_y,n_z)$, with $n_{x,y,z} \in\mathbb
{Z}$.  Modes with $\bn \neq \mathbf{0}$, being not supported by the grid, are observed at lower, aliased, modes, see Fig.~\ref{fig:aliasing}. In Fourier space,  $N$-th order PT fields  are  given by convolutions involving up to $N$ linear  fields at all possible modes. Therefore, in order to avoid that the resulting Fourier modes of the composite fields exceed the Nyquist limit, we will low-pass filter the linear fields at the $N$-dependent scale $\Lambda$, given by the Orszatz rule \cite{orszag1971predictability, Taruya2022},
\begin{equation}
    k\le \Lambda=\frac{2}{N+1}k_{\rm Nyq}=\frac{2}{N+1}\frac{\pi N_g}{L}\,.
\end{equation}
In the  following we will apply a isotropic top-hat filter in Fourier space, and will indicate explicitely the $\Lambda$-dependence of the fields. In particular, for the linear fields, $\vp_\Lambda(\bx)$ will indicate the Fourier transform of 
\begin{equation}
\vp_\Lambda(\bk)= \vp(\bk)\,\Theta\left(\Lambda-k\right)\,,
\label{eq:phicut}
\end{equation}
while, $\delta^{(n)}_{\Lambda}(\bx)$ and $\theta^{(n)}_{\Lambda}(\bx)$, for $n>1$, will indicate $n$-th order fields composed by $n$ linear fields  $\vp_\Lambda(\bx)$ .
%\MP{Meaning of $\Theta$ is tophat function?}

 Even ignoring aliasing, a cutoff scale is ultimately provided by the grid step, that is, the Nyquist frequency. Therefore, in field-theoretical language, the theory on the grid is unavoidably {\it regularized} by the presence of an unphysical cut-off, and the cut-off dependence must be carefully taken into account.  One needs to understand how $\Lambda$-independent  quantities can be defined from $\Lambda$-dependent fields and couplings, and how these quantities can be related to physical observables. That is, one needs to discuss the {\it renormalization} of the theory. Moreover, on a practical point of view, controlling the cut-off dependence of the theory allows using lower cut-off values, and then smaller number of grid points, increasing numerical efficiency. 
\subsection{Integrating out UV modes}

\begin{figure}
    \centering
\includegraphics[width=\textwidth]{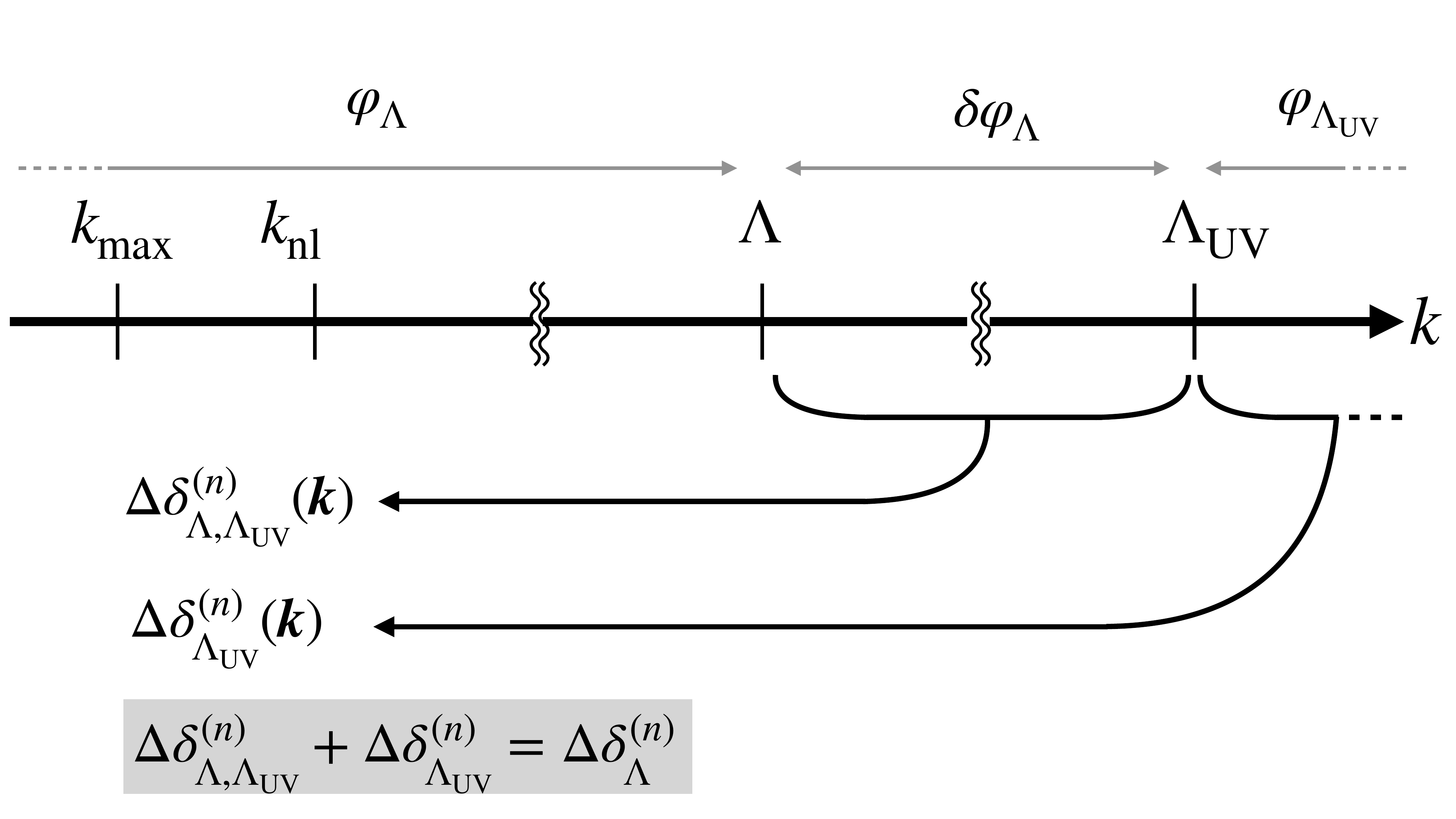}
    \caption{Schematic illustration of the relevant scales and fields defined in the paper. The theory is defined at a scale $\Lambda_{\rm uv} \gg k_{\rm nl}\gg k_{\rm max}$, as the sum of the PT contribution, Eq.~\eqref{eq:sPTuv}, and the UV counterterms, Eq.~\eqref{eq:uvcount}. By considering a scale $\L<\L_{\rm uv}$, we can define a new theory, again as the sum of a PT part (expressed in terms of the new IR fields $\vp_\L$) and UV counterterms. The latter are given by the sum of the UV counterterms of the original theory and the new contributions of Eq.~\eqref{eq:Deltadeltadef}, obtained by integrating out the $\delta \vp_\L$ fields.  }
    \label{fig:cutoffscales}
\end{figure}
In a Wilsonian approach \cite{Wilson:1974mb, Polchinski:1983gv}, the field theory is initially defined at a UV scale, $\Lambda_{\rm uv}$, and then related to theories  at lower cutoff scales $\Lambda$ by integrating out fluctuations of wavenumbers between $\Lambda$ and $\Lambda_{\rm uv}$. In the present cosmological context,  we assume the hierarchy (see Fig.~\ref{fig:cutoffscales}), 
\beq
k_{\rm max}\ll k_{\rm nl}\ll \Lambda_{\rm uv} \,,
\label{eq:hierarchy}
\eeq
where $k_{\rm max}$ is the largest mode for which we will compute the density field, while $k_{\rm nl}$ is the mode above which the fluid description on which PT is based breaks down. 

The UV theory is based on the EFTofLSS \cite{Baumann:2010tm, Pietroni:2011iz, Carrasco:2012cv}. We will consider PT models at $N$-th order, which in Fourier space, and omitting the time dependence, can be expressed as the sum of two types of contributions, 
\beq
\delta^{[N]}(\bk)= \delta^{[N], {\rm PT}}_{\Lambda_{\rm uv}}(\bk)+\Delta \delta^{[N]}_{\Lambda_{\rm uv}}(\bk) \,.
\label{eq:deltauv}
\eeq
The first term is the standard PT series truncated at order $N$,
\beq
\delta^{[N], {\rm PT}}_{\Lambda_{\rm uv}}(\bk) \equiv \sum_{n=1}^N \delta^{(n)}_{\Lambda_{\rm uv}}(\bk)\,,
\label{eq:sPTuv}
\eeq
with the $n$-th order contribution given by the convolution,
\beq
\delta^{(n)}_{\Lambda_{\rm uv}}(\bk)=
{\cal I}_{\bk;\bq_1,\cdots,\bq_n}
F_n(\bq_1,\cdots,\bq_n) \vp_{\Lambda_{\rm uv}}(\bq_1)\cdots\vp_{\Lambda_{\rm uv}}(\bq_n)\,.
\label{eq:deltanpt}
\eeq
The $F_n$'s are the usual symmetrized PT kernels \cite{Bernardeau_2002} (given at $n=2$ in Eq.~\eqref{eq:delta2}) and the filtered linear fields $\vp_{\Lambda_{\rm uv}}(\bq)$ have support only for wavenumbers $q\le \Lambda_{\rm uv}$, see Eq.~\eqref{eq:phicut}.
The second contribution in Eq.~\eqref{eq:deltauv} sums up the UV {\it counterterms} up to order $n=N$,
\beq\Delta \delta^{[N]}_{\Lambda_{\rm uv}}(\bk)\equiv \sum_{n=1}^N \Delta\delta^{(n)}_{\Lambda_{\rm uv}}(\bk)\,.
\label{eq:uvcount}
\eeq
The counterterms are built with the same fields $\vp_{\Lambda_{\rm uv}}$, and feed the effect of ``UV modes'' with $q>\Lambda_{\rm uv}$ on the modes $k$, in such a way that the PT expansion is well behaved and converges to the truth (e.g. a high resolution N-body simulation or, eventually, observational data). Because PT fails for scales $q>k_{\rm nl}$, the convergence is achieved up to corrections $O(k^2/k_{\rm nl}^2)$, whose coefficients are not computable in PT but must be fixed via {\it renormalization conditions} (see sect.~\ref{sect:runren}).
The quadratic decoupling of the nonperturbative corrections is a consequence of momentum conservation. Thanks to the hierarchy \eqref{eq:hierarchy}, cutoff dependent contributions of $O(k^2/\L_{\rm uv}^2)\ll O(k^2/k_{\rm nl}^2)$ are negligible in the UV theory.

We then consider a lower value for the cutoff, $\Lambda < \Lambda_{\rm uv}$ as shown in Fig.~\ref{fig:cutoffscales}, and derive the model, defined at the new scale $\L$, such that the full field is $\L$-independent, that is, imposing the condition,
\beq
 \delta^{[N],{\rm PT}}_\Lambda(\bk)+\Delta \delta^{[N]}_{\Lambda}(\bk)= \delta^{[N],{\rm PT}}_{\Lambda_{\rm uv}}(\bk)+\Delta \delta^{[N]}_{\Lambda_{\rm uv}}(\bk)=\delta^{[N]}(\bk) \,.
 \label{eq:Lindep}
\eeq
We split the UV field as,
\beq
\vp_{\Lambda_{\rm uv}}(\bq)=\vp_{\Lambda}(\bq)+\delta\vp_\Lambda(\bq)\,,
\label{eq:vpsplit}
\eeq
where $\delta\vp_\Lambda(\bq)$ has support on $\Lambda< q <\Lambda_{\rm uv}$,
\begin{align}
\vp_{\Lambda}(\bq)&=\vp_{\Lambda_{\rm uv}}(\bq)\Theta(\Lambda-q)=\vp(\bq)\Theta(\Lambda-q)\,,\nonumber\\
\delta\vp_\Lambda(\bq)&=\vp_{\Lambda_{\rm uv}}(\bq)\Theta(q-\Lambda)=\vp(\bq)\Theta(\Lambda_{\rm uv}-q)\Theta(q-\Lambda)\,.
\end{align}
The model at the scale $\L$ will be expressed in terms of the fields $\vp_\L(\bq)$, whereas the new UV fields $\delta\vp_\L(\bq)$ will be ``integrated out'', see below.
Implementing Eq.~\eqref{eq:vpsplit} in Eq.~\eqref{eq:deltanpt} gives
\beq
\delta^{(n)}_{\Lambda_{\rm uv}}(\bk)=\delta^{(n)}_{\Lambda}(\bk)+\Delta\delta^{(n)}_{\Lambda,\Lambda_{\rm uv}}(\bk)\,,
\label{eq:Deltadelta}
\eeq
where the first term is the standard PT contribution expressed in terms of the fields $\vp_\L(\bq)$,
\beq
\delta^{(n)}_{\Lambda}(\bk)=
{\cal I}_{\bk;\bq_1,\cdots,\bq_n}
F_n(\bq_1,\cdots,\bq_n) \vp_{\Lambda} (\bq_1)\cdots\vp_{\Lambda}(\bq_n)\,,
\label{eq:deltanptL}
\eeq
while the second term contains the couplings between IR and UV fields,
\beq
\Delta\delta^{(n)}_{\Lambda,\Lambda_{\rm uv}}(\bk)=\sum_{p=1}^n \left(\begin{array}{c}n\\ p\end{array}\right){\cal I}_{\bk;\bq_1,\cdots,\bq_n}F_n(\bq_1,\cdots,\bq_n) \delta\vp_{\Lambda}(\bq_1)\cdots \delta\vp_{\Lambda}(\bq_p)\vp_{\Lambda}(\bq_{p+1})\cdots\vp_{\Lambda}(\bq_n)\,.
\label{eq:Deltadeltadef}
\eeq
Defining the standard PT contribution in Eq.~\eqref{eq:Lindep} analogously to Eq.~\eqref{eq:sPTuv},
\beq
\delta^{[N], {\rm PT}}_{\Lambda}(\bk) \equiv \sum_{n=1}^N \delta^{(n)}_{\Lambda}(\bk)\,,
\label{eq:sPTL}
\eeq
we get the counterterms at the new scale $\L$,
\beq
\Delta\delta^{[N]}_{\Lambda}(\bk)=\Delta\delta^{[N]}_{\Lambda,\Lambda_{\rm uv}}(\bk)+\Delta\delta^{[N]}_{\Lambda_{\rm uv}}(\bk)\,,
\label{eq:ctL}
\eeq
where
\beq
\Delta\delta^{[N]}_{\Lambda,\Lambda_{\rm uv}}(\bk) \equiv \sum_{n=1}^N\Delta\delta^{(n)}_{\Lambda,\Lambda_{\rm uv}}(\bk).
\eeq
It is important to notice the difference between the two contributions in Eq.~\eqref{eq:ctL} (see also Fig.~\ref{fig:cutoffscales}). The second term, $\Delta\delta^{[N]}_{\Lambda_{\rm uv}}(\bk)$, is not computable in PT, but can only be fixed by imposing renormalization conditions, obtained by measurements on simulations or on real data. On the other hand, the second term vanishes for $\L=\L_{\rm uv}$ and can be computed in PT at any $\L$, according to Eq.~\eqref{eq:Deltadeltadef}. 
\subsection{Structure of the counterterms}
In order to understand the structure of the counterterms, we can inspect in details the lowest orders, from the definition \eqref{eq:Deltadeltadef}.
At $n=1$ we have 
\beq
\Delta\delta^{(1)}_{\Lambda,\Lambda_{\rm uv}}(\bk)= \int d^3 q \,\delta_D(\bk-\bq)  \delta\vp_{\Lambda}(\bq)=\delta\vp_{\Lambda}(\bk)=0\,,
\eeq
which vanishes because $k < \Lambda$. For the same reason, at $n=2$ we have just the contribution involving two UV fields,
\beq
\Delta\delta^{(2)}_{\Lambda,\Lambda_{\rm uv}}(\bk)= {\cal I}_{\bk;\bq_1,\bq_2}F_2(\bq_1,\bq_2) \delta\vp_{\Lambda}(\bq_1) \delta\vp_{\Lambda}(\bq_2)\,.
\eeq
Only the IR fields $\vp_\L(\bq)$  are described deterministically by the PT model at the scale $\Lambda$. On the other hand, the $\delta\vp_\L(\bq)$ fields are integrated out, that is, we take the expectation values of their products. This gives,
\beq
\langle \Delta\delta^{(2)}_{\Lambda,\Lambda_{\rm uv}}(\bk)\rangle = {\cal I}_{\bk;\bq_1,\bq_2}F_2(\bq_1,\bq_2) \langle\delta\vp_{\Lambda}(\bq_1) \delta\vp_{\Lambda}(\bq_2)\rangle=\int \frac{d^3 q}{(2\pi)^3}F_2(\bq,-\bq)P_{\Lambda,\Lambda_{\rm uv}}(q)=0\,,
\label{eq:van2}
\eeq
where we have defined $P_{\Lambda,\Lambda_{\rm uv}}(q)$ as
\beq
\langle\delta\vp_{\Lambda}(\bq_1) \delta\vp_{\Lambda}(\bq_2)\rangle \equiv (2\pi)^3\delta_D(\bq_1+\bq_2)P_{\Lambda,\Lambda_{\rm uv}}(q_2)\,,
\eeq
that is,
\beq
P_{\Lambda,\Lambda_{\rm uv}}(q)\equiv P(q)\Theta(\Lambda_{\rm uv}-q)^2\Theta(q-\Lambda)^2\,.
\eeq
 The vanishing of Eq.~\eqref{eq:van2} comes from momentum conservation, which implies the following decoupling behavior of the second order kernel,
\beq
F_2\left(\frac{\bk}{2}+\bq,\frac{\bk}{2}-\bq\right) = O\left(\frac{k^2}{q^2}\right)\,,\qquad {\rm for}\; k\ll q\,.
\eeq
Moreover, for Gaussian initial conditions, the stochastic field is uncorrelated to the linear field,
\beq
\langle\Delta\delta^{(2)}_{\Lambda,\Lambda_{\rm uv}}(\bk) \vp_\Lambda(\bk')\rangle ={\cal I}_{\bk;\bq_1,\bq_2}F_2(\bq_1,\bq_2) \langle\delta\vp_{\Lambda}(\bq_1) \delta\vp_{\Lambda}(\bq_2)\vp_\Lambda(\bk')\rangle   =0,
\eeq
and has a power spectrum given by,
\beq
\langle\Delta\delta^{(2)}_{\Lambda,\Lambda_{\rm uv}}(\bk) \Delta\delta^{(2)}_{\Lambda,\Lambda_{\rm uv}}(-\bk)\rangle' = P^{22}_{\Lambda,\Lambda_{\rm uv}}(k)\,,
\eeq
where the prime indicates that the factor $(2\pi)^3\delta_D(0)$ has been omitted from the average, and,
\beq
P^{22}_{\Lambda,\Lambda_{\rm uv}}(k)\equiv 
2 \int\frac{d^3 q}{(2 \pi)^3}F_2(\bq, \bk-\bq)^2 P_{\Lambda,\Lambda_{\rm uv}}(q)P_{\Lambda,\Lambda_{\rm uv}}(|\bk-\bq|)\simeq  \frac{127}{294}k^4  \int \frac{d^3 q}{(2 \pi)^3} \left(\frac{P_{\Lambda,\Lambda_{\rm uv}}(q)}{q^2}\right)^2  \,,
\eeq
with the last equality holding in the $k\ll \Lambda$ regime. 

In summary, at lowest order in $k/\Lambda$, we can write the second order counterterm field as,
\beq
\Delta\delta^{(2)}_{\Lambda,\Lambda_{\rm uv}}(\bk)\simeq \frac{k^2}{k^2_{\rm nl}} \,\epsilon^{(2)}_{\Lambda,\Lambda_{\rm uv}},
\label{eq:eps2}
\eeq
where the stochastic coefficient $\epsilon^{(2)}_{\Lambda,\Lambda_{\rm uv}}$ is uncorrelated to the linear field and has constant variance, given by
\beq
P^\epsilon_{\Lambda,\Lambda_{\rm uv}}= \lim_{k/\L\to 0}\frac{k^4_{\rm nl}}{k^4} P^{22}_{\Lambda,\Lambda_{\rm uv}}(k)\,.
\eeq
All its higher order cumulants can be obtained by analogous computations in PT.

At order $n=3$ two new contributions arise,
\begin{align}
\Delta\delta^{(3)}_{\Lambda,\Lambda_{\rm uv}}(\bk)&= \Delta\delta^{(3),a}_{\Lambda,\Lambda_{\rm uv}}(\bk)+\Delta\delta^{(3),b}_{\Lambda,\Lambda_{\rm uv}}(\bk)\,,
\end{align}
with
\begin{align}
\Delta\delta^{(3),a}_{\Lambda,\Lambda_{\rm uv}}(\bk)&\equiv 3 \,{\cal I}_{\bk;\bq_1,\bq_2,\bq_3}F_3(\bq_1,\bq_2,\bq_3) \delta\vp_{\Lambda}(\bq_1)\delta\vp_{\Lambda}(\bq_2)\vp_{\Lambda}(\bq_3)\,,\\
\Delta\delta^{(3),b}_{\Lambda,\Lambda_{\rm uv}}(\bk)&\equiv \,{\cal I}_{\bk;\bq_1,\bq_2,\bq_3}F_3(\bq_1,\bq_2,\bq_3) \delta\vp_{\Lambda}(\bq_1)\delta\vp_{\Lambda}(\bq_2)\delta\vp_{\Lambda}(\bq_3)\,,\label{eq:delta3b}
\end{align}
whereas the contribution with only one $\delta\vp_{\Lambda}$ vanishes, again, by momentum conservation.
The stochastic field $\Delta\delta^{(3),a}_{\Lambda,\Lambda_{\rm uv}}(\bk)$ has vanishing expectation value and is correlated with the linear field,
\begin{align}  
&\langle \Delta\delta^{(3),a}_{\Lambda,\Lambda_{\rm uv}}(\bk)\rangle =0\,,\nonumber\\
&\langle \Delta\delta^{(3),a}_{\Lambda,\Lambda_{\rm uv}}(\bk) \vp_\Lambda(-\bk)\rangle'  = 3 P_\Lambda(k) \int \frac{d^3 q}{(2\pi)^3} F_3(\bk,\bq,-\bq)  P_{\Lambda,\Lambda_{\rm uv}}(q)\,,\nonumber\\
&\qquad\qquad= - \left(\frac{61}{630} \int \frac{d^3q}{(2 \pi)^3} \frac{P_{\Lambda,\Lambda_{\rm uv}}(q)}{q^2} \right)\, k^2 \left(1+O\left(\frac{k^2}{q^2}\right)\right)P_\Lambda(k)\simeq - \,c^{(3)}_{\Lambda,\Lambda_{\rm uv}} \frac{k^2}{k^2_{\rm nl}} P_\Lambda(k)\,,
\label{eq:c2def}
\end{align}
where, again, the last line corresponds to the $k\ll \Lambda$ limit,  $P_\Lambda(k)$ is the IR  linear power spectrum,
\beq
P_\Lambda(k)\equiv\langle\vp_\Lambda(\bq)\vp_\Lambda(-\bq)\rangle'\,,
\eeq
and we have defined the  coefficient 
\beq
c^{(3)}_{\Lambda,\Lambda_{\rm uv}}\equiv \frac{61}{630} \int \frac{d^3q}{(2 \pi)^3} \frac{P_{\Lambda,\Lambda_{\rm uv}}(q)}{q^2} \,,
\label{eq:cs3}
\eeq
which gives the perturbative contribution at third order to the speed of sound of the EFTofLSS \cite{Carrasco:2012cv}.
The power spectrum of the $\Delta\delta^{(3),a}_{\Lambda,\Lambda_{\rm uv}}(\bk)$ field is given by
\begin{align}
&\langle \Delta\delta^{(3),a}_{\Lambda,\Lambda_{\rm uv}}(\bk) \Delta\delta^{(3),a}_{\Lambda,\Lambda_{\rm uv}}(-\bk)  \rangle'=  P_\Lambda(k) \left(3\int \frac{d^3 q}{(2 \pi)^3}  F_3(\bk,\bq,-\bq) P_{\Lambda,\Lambda_{\rm uv}}(q)\right)^2  \nonumber\\
& +2 \int \frac{d^3 q}{(2 \pi)^3}\frac{d^3 p}{(2 \pi)^3} \left[F_3(\bk-\bq+\bp,\bq,-\bp) \right]^2 P_\Lambda(|\bk-\bq+\bp|) P_{\Lambda,\Lambda_{\rm uv}}(q)P_{\Lambda,\Lambda_{\rm uv}}(p)\nonumber\\
&\simeq \left[\left(c^{(3)}_{\Lambda,\Lambda_{\rm uv}}\right)^2 + \sigma^{(3)}_{\Lambda,\Lambda_{\rm uv}} \right] \frac{k^4}{k^4_{\rm nl}}  P_\Lambda(k) +O\left(\frac{k^6}{k^6_{\rm nl}}\right),
\label{eq:Delta3a}
\end{align}
In other terms, we can write,
\beq
\Delta\delta^{(3),a}_{\Lambda,\Lambda_{\rm uv}}(\bk)\simeq - \left(c^{(3)}_{\Lambda,\Lambda_{\rm uv}} + \eta^{(3)}_{\Lambda,\Lambda_{\rm uv}}\right) \frac{k^2}{k^2_{\rm nl}} \vp_\Lambda(\bk)\,,
\label{eq:cs3b}
\eeq
where $c^{(3)}_{\Lambda,\Lambda_{\rm uv}}$ is defined in \eqref{eq:cs3}, and the stochastic coefficient $\eta^{(3)}_{\Lambda,\Lambda_{\rm uv}}$ is scale independent, has vanishing expectation value and variance $\sigma^{(3)}_{\Lambda,\Lambda_{\rm uv}}$, which can be read from the small $k$ limit of the second contribution in Eq.~\eqref{eq:Delta3a}.

The term $\Delta\delta^{(3),b}_{\Lambda,\Lambda_{\rm uv}}(\bk)$, at small $k$, gives a contribution similar to that  obtained at second order, Eq.~\eqref{eq:eps2},
\beq
\Delta\delta^{(3),b}_{\Lambda,\Lambda_{\rm uv}}(\bk)\simeq \frac{k^2}{k^2_{\rm nl}} \,\epsilon^{(3)}_{\Lambda,\Lambda_{\rm uv}}\,,
\label{eq:eps3}
\eeq
where $\epsilon^{(3)}_{\Lambda,\Lambda_{\rm uv}}$ has zero average, is uncorrelated to the linear field, and has a power spectrum given by the two-loop UV integral obtained by squaring Eq.~\eqref{eq:delta3b}. 

\subsection{Quadratic higher derivative terms}
\label{sub:highder}
At $n>3$, counterterms of the type \eqref{eq:Deltadeltadef} with $n-p=2$ contain contributions quadratic in the IR linear fields $\vp_\Lambda$. These terms correlate with $\vp_\gamma^{(2)}$, and therefore enter in the determination of the bootstrap coefficient in Eq.~\eqref{eq:delta2boot}, as we will discuss in Sect.~\ref{sub:estim}. In order to extract $\varepsilon_\gamma$ reliably, these counterterms should be taken into account. In particular, at $n=4$, integrating out two linear fields $\delta\vp_\Lambda(\bx)$ one obtains contributions of the following form, in configuration space,
\begin{align}
& c^{(n)}_{\vp^2;\Lambda,\Lambda_{\rm uv}} \,\frac{\partial^2}{k_{\rm nl}^2}\vp_\Lambda^2(\bx)\,,\quad c^{(n)}_{\vp_\beta;\Lambda,\Lambda_{\rm uv}}\, \frac{\partial^2}{k_{\rm nl}^2}\vp^{(2)}_{\beta,\Lambda}(\bx)\,,\nonumber \\
& c^{(n)}_{\vp_\gamma;\Lambda,\Lambda_{\rm uv}} \,\frac{\partial^2}{k_{\rm nl}^2}\vp^{(2)}_{\gamma,\Lambda}(\bx)
\,,\quad
c^{(n)}_{\tilde\gamma;\Lambda,\Lambda_{\rm uv}}\, \frac{\partial^2}{k_{\rm nl}^2}\,\vp^{(2)}_{\tilde\gamma,\Lambda}(\bx)\,.
\label{eq:highder}
\end{align}
The quadratic operator  $\vp_\Lambda^2(\bx)$ is just the square of $\vp_\Lambda(\bx)$, $\vp^{(2)}_{\beta,\Lambda}(\bx)$ and $\vp^{(2)}_{\gamma,\Lambda}(\bx)$ have been defined in Eqs.~\eqref{eq:phibeta} and \eqref{eq:phigamma}, respectively, while $\vp^{(2)}_{\tilde\gamma,\Lambda}(\bx)$ is the Fourier transform of
\beq
\vp^{(2)}_{\tilde\gamma,\Lambda}(\bk)\equiv {\cal I}_{\bk;\bp_1,\bp_2}\, \tilde \gamma(\bp_1,\bp_2)\,\vp_{\Lambda}(\bp_1)\vp_{\Lambda}(\bp_2)\,,
\eeq
with
\beq
\tilde\gamma(\bp_1,\bp_2)\equiv \frac{\bp_1\cdot\bp_2}{|\bp_1+\bp_2|^2}\gamma(\bp_1,\bp_2)\,.
\eeq
The $\Lambda$ dependence of the coefficients can be obtained by a perturbative computation, analogously to what discussed for the ``speed of sound'' coefficient, $c^{(3)}_{\Lambda,\Lambda_{\rm uv}}$, see Eq.~\eqref{eq:cs3}. At fourth order, we get,
\beq
c^{(4)}_{X;\Lambda,\Lambda_{\rm uv}}=\frac{2096}{33957 \,\pi^2} \, C_X  \,k_{\rm nl}^2 \int^{\Lambda_{\rm uv}}_\Lambda P(\Lambda') d\Lambda'\,,
\label{eq:runningchd}
\eeq
where $X=\{\vp^2,\,\vp_\gamma,\,\vp_\beta,\,\vp_{\tilde\gamma}\}$, 
and
\beq C_{\vp^2}=1\,,\quad C_{\vp_\gamma}=\frac{4223}{10480}\,,\quad C_{\vp_\beta}=\frac{32879}{20960}\,,\quad C_{\tilde\gamma}=\frac{1503}{1310}\,.
\eeq
All these terms are ``higher derivatives'', that is, suppressed by $O(k^2/k_{\rm nl}^2)$ in Fourier space. The absence of contributions $O(1)$ is again a consequence of momentum conservation. 
\subsection{Running and renormalization}
\label{sect:runren}

 The $\Lambda$-dependence (the {\it running}) of the counterterms in $\Delta\delta^{[N]}_{\Lambda}(\bk)$ is  determined by that of $\Delta\delta^{[N]}_{\Lambda,\Lambda_{\rm uv}}(\bk)$, see Eq.~\eqref{eq:ctL},
 \beq
 \Lambda \frac{d}{d\Lambda} \Delta\delta^{[N]}_{\Lambda}(\bk)= \Lambda \frac{d}{d\Lambda} \Delta\delta^{[N]}_{\Lambda,\Lambda_{\rm uv}}(\bk)\,,
 \eeq
 and therefore, as we have discussed, it can be computed in PT.
 To fully define the model,  one has to impose renormalization conditions, that is, boundary conditions for the equations for the running of $\Delta\delta^{[N]}_{\Lambda}(\bk)$ at a given scale $\Lambda=\bar \Lambda$, and for a given external momentum $\bk=\bar \bk$. One possibility is to give the renormalization conditions at $\bar \Lambda=\Lambda_{\rm uv}$. Assuming the hierarchy \eqref{eq:hierarchy},
the form of the counterterms is fixed by the available symmetries, that is, mass and momentum conservation, rotational invariance and EGI, which, up to third order and at leading order in $k$, give
\begin{align}
\Delta\delta^{(2)}_{\Lambda_{\rm uv}}(\bk)& \simeq \frac{k^2}{k^2_{\rm nl}} \,\epsilon^{(2)}_{\Lambda_{\rm uv}}\,,\label{eq:delta2Lam}\\
\Delta\delta^{(3)}_{\Lambda_{\rm uv}}(\bk)&\simeq \frac{k^2}{k^2_{\rm nl}} \,\epsilon^{(3)}_{\Lambda_{\rm uv}} - \left(c^{(3)}_{\Lambda_{\rm uv}} + \eta^{(3)}_{\Lambda_{\rm uv}}\right) \frac{k^2}{k^2_{\rm nl}} \vp_{\Lambda_{\rm uv}}(\bk)+O\left(\frac{k^4}{k_{\rm nl}^4}\right)\,.\label{eq:delta3Lam}
\end{align}
Notice that these structures are the same as those for the pertrubative counterterms $\Delta\delta^{(2)}_{\Lambda,\Lambda_{\rm uv}}(\bk)$ and $\Delta\delta^{(3)}_{\Lambda,\Lambda_{\rm uv}}(\bk)$ from Eqs.~\eqref{eq:eps2}, \eqref{eq:cs3}, and \eqref{eq:eps3}. 
Symmetry arguments enforce the same expansion for the counterterms holds at any lower scale $\Lambda$, as long as the hierarchy $k\ll k_{\rm nl}\ll \Lambda$ holds,
\begin{align}
\Delta\delta^{(2)}_{\L}(\bk)& \simeq \frac{k^2}{k^2_{\rm nl}} \,\epsilon^{(2)}_{\L}\,,\label{eq:delta2Lam2}\\
\Delta\delta^{(3)}_{\L}(\bk)&\simeq \frac{k^2}{k^2_{\rm nl}} \,\epsilon^{(3)}_{\L} - \left(c^{(3)}_{\L} + \eta^{(3)}_{\L}\right) \frac{k^2}{k^2_{\rm nl}} \vp_{\L}(\bk)+O\left(\frac{k^4}{k_{\rm nl}^4}\right)\,.\label{eq:delta3Lam2}
\end{align}
Focusing on the parameter $c^{(3)}_{\Lambda}$, from Eq.~\eqref{eq:ctL}, we get
\beq
c^{[3]}_{\Lambda}=c^{(3)}_{\Lambda}=c^{(3)}_{\Lambda,\L_{\rm uv}}+c^{(3)}_{\L_{\rm uv}}.
\eeq
It runs with $\Lambda$ as (see Eq.~\eqref{eq:cs3}),
\beq
\Lambda \frac{d}{d\Lambda} c^{(3)}_{\Lambda}=\Lambda \frac{d}{d\Lambda}c^{(3)}_{\Lambda,\Lambda_{\rm uv}} = - \frac{61}{1260 \,\pi^2} \Lambda \,k_{\rm nl}^2 P(\Lambda)\,,
\label{eq:runningc}
\eeq
and therefore it can be expressed, at any $\Lambda$, as 
\beq
c^{(3)}_\Lambda= c^{(3)}_{\Lambda_{\rm uv}}+\frac{61}{1260 \,\pi^2}  \,k_{\rm nl}^2  \int_{\Lambda}^{\Lambda_{\rm uv}} d\Lambda^\prime P(\Lambda^\prime) \,,
\label{eq:runc3}
\eeq
where the UV boundary condition, $c^{(3)}_{\Lambda_{\rm uv}}$,  has to be fixed by the renormalization condition.

At $N=5$, the sound speed coefficient gets a new contribution, 
\beq
\Delta\delta^{[5]}_{\L}(\bk)= \Delta\delta^{(3)}_{\L}(\bk)+\Delta\delta^{(5)}_{\L}(\bk) =  - c^{[5]}_{\L} \frac{k^2}{k^2_{\rm nl}} \vp_{\L}(\bk)+\cdots\,,
\eeq
where we have defined
\begin{align}
c^{[5]}_{\Lambda}&\equiv  c^{(3)}_\Lambda+c^{(5)}_\Lambda\,.
\end{align}
The running of $c^{(5)}_\Lambda$, obtained analogously to \eqref{eq:runningc}, is given by the two-loop contribution,
\begin{align} 
 c^{(5)}_{\Lambda}&= c^{(5)}_{\Lambda_{\rm uv}}-\int_\L^{\L_{\rm uv}} d\L'\,\left(\frac{d}{d \L'}c_{\L'}^{(5)}\right)\,,
 \nonumber\\
 &=
 c^{(5)}_{\Lambda_{\rm uv}} 
 +15 \,\lim_{\bk \to 0} \frac{k^2_{\rm nl}}{k^2}  \int \frac{d^3 q}{(2\pi)^3}\frac{d^3 p} {(2\pi)^3} F_5(\bk,\bq,-\bq,\bp,-\bp)P_{\Lambda,\Lambda_{\rm uv}}(q)P_{\Lambda,\Lambda_{\rm uv}}(p)\,.
\label{eq:runningc5}
\end{align}

A possible, physically motivated, renormalization condition is to require that the zero momentum limit of the cross-correlator between  the $N$-th order field (after the subtraction of the linear contribution) and the linear field, 
\beq
 \lim_{\bk\to 0} \left(\frac{\langle  \delta^{[N]}(\bk) \vp_\Lambda(-\bk)\rangle'}{ P_\Lambda(k)}-1\right)\frac{k_{\rm nl}^2}{k^2}=\lim_{\Lambda\to 0}\lim_{\bk\to 0} \frac{\langle  \Delta\delta^{[N]}_\Lambda(\bk) \vp_\Lambda(-\bk)\rangle'}{ P_\Lambda(k)}\frac{k_{\rm nl}^2}{k^2}=-c^{[N]}_{\Lambda=0}\,,
 %\label{eq:cl0}
\eeq
coincides with the measured value for every $N$,
\beq
 \bar c\equiv -\lim_{\bk\to 0} \left(\frac{\langle  \bar\delta(\bk) \vp(-\bk)\rangle'}{ P(k)}-1\right)\frac{k_{\rm nl}^2}{k^2}\,,
 \label{eq:cl0}
\eeq
where $\bar\delta(\bk)$ is the data field (e.g. from N-body simulations).
According to Eqs.~\eqref{eq:runc3} and \eqref{eq:runningc5}  imposing $c^{[3]}_{\Lambda=0}=c^{[5]}_{\Lambda=0}=\bar c$,
gives the following relation,
\beq
c^{[3]}_{\Lambda} -c^{[5]}_{\Lambda}=  -\int_{0}^{\Lambda} d\Lambda^\prime \left(\frac{d}{d\Lambda^\prime} c^{(5)}_{\Lambda^\prime}\right)\,,
\label{eq:c3c5uv}
\eeq
where the difference at the RHS is given by the two-loop integral in  Eq.~\eqref{eq:runningc5}. 

On the other hand, the purely non-perturbative contributions to the sound speed would formally correspond to the $N\to \infty$ limit of the coefficient in the $\Lambda\to \infty$ limit,
\beq
c_s^2= \lim_{N\to \infty} \,\lim_{\Lambda\to \infty} c^{[N]}_{\Lambda}\,.
\eeq

By taking the derivative of Eq.~\eqref{eq:runningchd} with respect to $\L$ we obtain the running of the coefficients of the higher derivative operators,
\beq
\L \frac{d}{d \L}c^{(4)}_{X;\Lambda,\Lambda_{\rm uv}}=- \frac{2096}{33957 \,\pi^2} \, C_X  \,\L\,k_{\rm nl}^2  P(\Lambda) \,.
\label{eq:runcXX}
\eeq

\begin{figure}
    \centering
    \includegraphics[width=\textwidth]{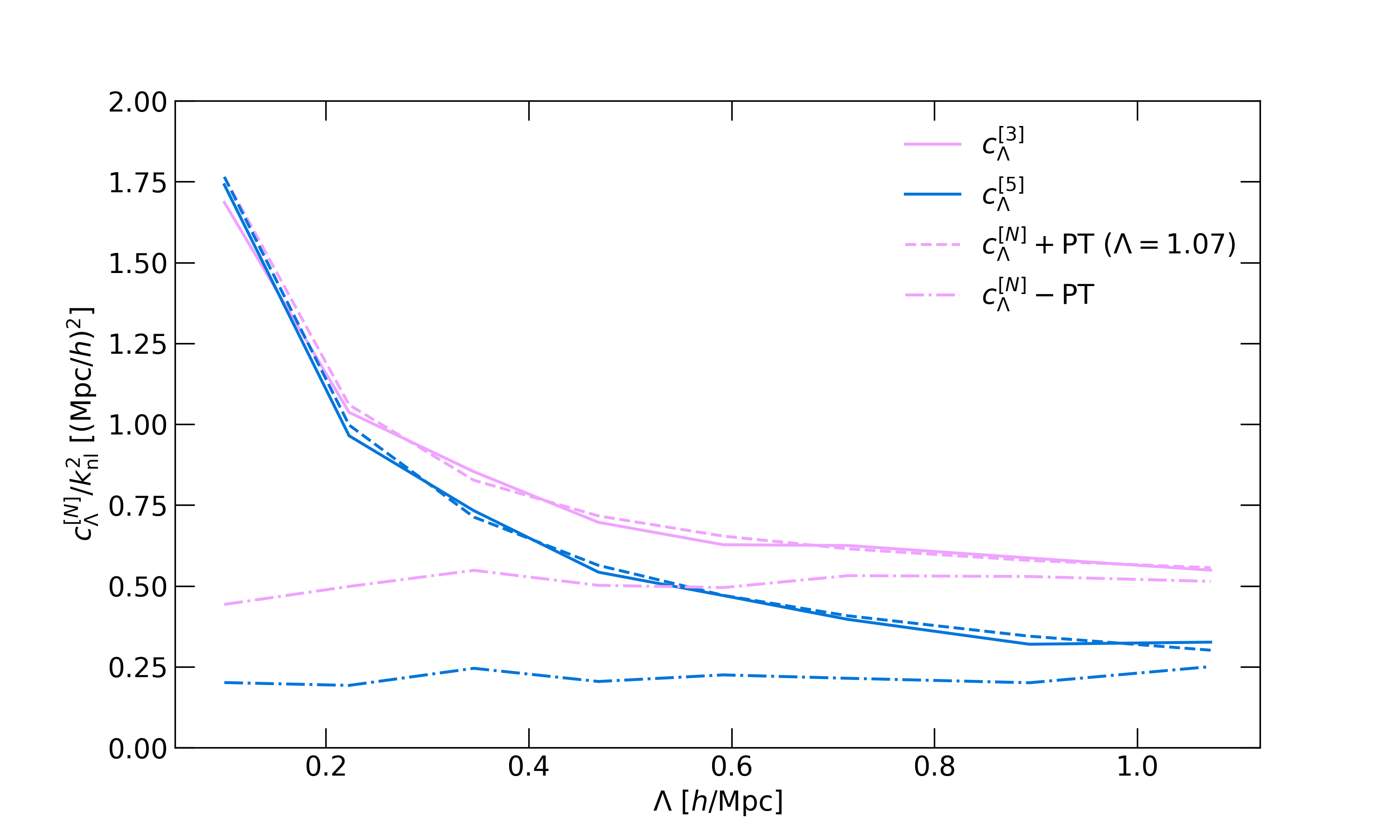}
    \caption{Running of $c^{[3]}_{\Lambda}$ (blue lines) and $c^{[5]}_{\Lambda}$ (pink lines) as measured from $N$-body simulations (solid lines). The running predicted by PT from equations \eqref{eq:runc3} and \eqref{eq:runningc5} is shown by the dashed lines. To ease the comparison, we shifted the the PT curves in such a way as to match the N-body ones at $\L=\L_{\rm uv}=1.07 \;\hMpc$. The difference between the N-body measurement and the (not shifted) PT computation is given by the dot-dashed lines. Their $\L$-independent values approximate the UV quantities $c^{[3]}_{\Lambda_{\rm uv}}$ and $c^{[5]}_{\Lambda_{\rm uv}}$.}
    \label{fig:Runcs}
\end{figure}

\section{Setting up the analysis}
\label{sect:Likelihood}
\subsection{Likelihood and data}
In the following, we will compare the bootstrap-PT model with $\L$CDM simulations, and will test the reliability of our setup in extracting the non-linear bootstrap parameter $\varepsilon_\gamma$, as a first step towards more realistic searches for physics beyond $\L$CDM from real data. We will assume a Gaussian likelihood of the form, 
\beq
\log P_N[\delta| \delta_{in}](k_{\rm max};\Lambda, \{\alpha^i_\Lambda\})=
-\frac{1}{2}\sum_{\bf n}^{n_{\rm max}} \left[ \frac{\left|\delta(\bk_{\bf n})-\delta^{[N]}(\bk_{\bf n})\right|^2}{L^3 \,p_\epsilon(k_{\bf n})}\right]+{\rm const}\,,
\label{eq:loglikF}
\eeq
where $\bk_{\bf n}= \frac{2 \pi}{L}{\bf n}$, $n_{\rm max}= k_{\rm max}\, L/(2\pi)$  and we assume constant noise  
\beq
p_\epsilon(k_{\bf n})=\frac{1}{\bar n}\,,
\eeq
where $\bar n$ is the number density of sources. We indicate with $ \{\alpha^i_\Lambda\}$ the set of parameters of the model $\delta^{[N]}(\bk_{\bf n})$, that is, $\{c^{[N]}_\Lambda,\varepsilon_\gamma,\cdots\}$, where dots may include stochastic and higher derivative terms.  
The data $\delta(\bk_{\bf n})$ are computed from the overdensity field of dark matter particles measured from a $N$-body simulation with the same initial conditions $\vp(\bk)$, used in the computation of the model $\delta^{[N]}(\bk_{\bf n})$ from \texttt{GridSPT}. That is, both the $N$-body simulations and the \texttt{GridSPT} codes share the same initial random seeds and adopt identical cosmological parameters. 

We use a cosmological $N$-body simulation performed as a part of the ongoing \texttt{Dark Quest II} project (Nishimichi \textit{et al.} in prep.). The simulation employs $3\,000^3$ particles in a comoving volume of $(1\,h^{-1}\mathrm{Gpc})^3$. The cosmological model assumed in the simulation (and \texttt{GridSPT}) is the best-fitting flat-geometry $\Lambda$CDM cosmology to the temperature and polarization spectra measured from Planck satellite~\cite{Planck_2015}. However, we ignore massive neutrinos, for which $\sum m_\nu = 0.06\,\mathrm{eV}$ is assumed in the reference, for simplicity in comparing with the perturbation-theory predictions.\footnote{The Dark Quest II. simulation suits treats massive neutrinos using the linear response approach~\cite{Ali_Haimoud_2012}. Massive neutrinos lead to scale-dependent linear growth, which can be implemented consistently in perturbation theory. We opt not to include this and consider a simpler case without scale-dependent linear growth in this paper.}

The simulation particles are initialized using the second-order Lagrangian perturbation theory \cite{Scoccimarro:1997gr,Crocce:2006ve} at $z=91$, sourced by a Gaussian random field following the power spectrum computed by the \texttt{CLASS} Boltzmann solver~\cite{Blas:2011rf}. The particle positions and velocities are then evolved to $z=0$ using \texttt{GINKAKU} Tree-Partcle Mesh code, developed based on the FDPS library~\cite{iwasawa16,Namekata_2018}.\footnote{https://github.com/FDPS/FDPS} We use the Particle-Mesh extension in FDPS from the \texttt{GreeM} code \citep{Yoshikawa_2005,Ishiyama_2009,Ishiyama_2012}. The tree force calculation is accelerated by the \texttt{Phantom-GRAPE} library~\citep{Tanikawa_2012,Tanikawa_2013} with the AVX-512 instructions~\citep{Yoshikawa_2018}.

We analyze the snapshot stored at $z=1$. The matter density field is constructed by cloud-in-cells (CIC) density assignment scheme \cite{hockney81} at $2\,000^3$ grid points, corresponding to a $0.5\,h^{-1}\mathrm{Mpc}$ resolution.

We artificially add noise to the dark matter density field to mimic a realistic catalog of point-like sources such as galaxies. The procedure is detailed in appendix \ref{app:noise}, and consists of:
\begin{enumerate}
    \item assuming a value for the average density of sources $\bar{n}=2\cdot10^{-3}\,(\hMpc)^3$, which means $2$ million points per cubic comoving $h^{-1}\,{\rm Gpc}$, to mimic a Euclid-like survey \cite{Amendola:2016saw};
    \item computing the number count of points per grid cell as $\tilde{N}(\bx)=(1+\delta_m(\bx))\bar{n}\,\Delta x^3$, where $\delta_m(\bx)$ is the dark matter density field and $\Delta x\equiv L/N_g$ is the size of a grid cell;
    \item sampling a new number count for each cell $N(\bx)$ by assuming $\tilde{N}(\bx)$ as the average of a Poisson distribution, $N(\bx)\sim{
\rm Poisson}(\tilde{N}(\bx))$;
    \item recalculating the noisy data by applying the definition $\delta(\bx)=\frac{N(\bx)}{\bar{N}}-1$, where $\bar{N}\equiv L^3\bar{n}$ is the average value of $N(\bx)$.
\end{enumerate}

 We generate ten independent realizations of the noisy data in this way, which will be all analyzed in the next sections.

\subsection{Models}
We focus our analysis on two PT models, $N=3$ and $N=5$. In order to check the proper  renormalization of the models we  consider different values of $\L$ and $k_{\rm max}$. Based on the discussion in Sect.~\ref{sect:LambdaDep}, the renormalized model for $N=3$ is
\begin{equation}
    \delta^{[3]}(\bk)= \delta^{[3],{\rm PT}}_\L(\bk)+ \frac{1}{2} \varepsilon_\gamma \,a^{(2)}_{\rm EdS} \,\vp^{(2)}_{\gamma,\Lambda}(\bx)+\Delta\delta^{[3]}_\L(\bk)\,,
\end{equation}
where $\delta^{[3],{\rm PT}}_\L(\bk)$ is computed in the EdS approximation. The set of counterterms can be red by summing Eqs.~\eqref{eq:delta2Lam2} and \eqref{eq:delta3Lam2}, and has the following form,
\begin{equation}
    \Delta\delta^{[3]}_\L(\bk)= -c^{[3]}_\L \frac{k^2}{k_{\rm nl}^2}\vp_\L(\bk) + \frac{k^2}{k_{\rm nl}^2}\left(\epsilon^{[3]}_\L- \eta^{[3]}_\L\,\vp_\L(\bk)\right)\,,
    \label{eq:ct3}
\end{equation}
up to $O(k^2/k^4_{\rm nl})$ corrections. Here, we defined  $\epsilon^{[3]}_\L\equiv\epsilon^{(2)}_\L+\epsilon^{(3)}_\L$, while $c^{[3]}_\L= c^{(3)}_\L$ and $\eta^{[3]}_\L=\eta^{(3)}_\L$\,.

The $N=5$ model is
\begin{equation}
    \delta^{[5]}(\bk)= \delta^{[5],{\rm PT}}_\L(\bk)+ \frac{1}{2} \varepsilon_\gamma \,a^{(2)}_{\rm EdS} \,\vp^{(2)}_{\gamma,\Lambda}(\bx)+\Delta\delta^{[5]}_\L(\bk)\,,
\end{equation}
with $\delta^{[5],{\rm PT}}_\L(\bk)$ also computed via \texttt{GridSPT} in the EdS limit, and
\begin{align}
     \Delta\delta^{[5]}_\L(\bk)=& -c^{[5]}_\L \frac{k^2}{k_{\rm nl}^2}\vp_\L(\bk) +\frac{k^2}{k_{\rm nl}^2}\sum_{X}c^{[5]}_{X,\L}  X_\L(\bk)\nonumber\\
     &+ \frac{k^2}{k_{\rm nl}^2}\left(\epsilon^{[5]}_\L- \eta^{[5]}_\L\,\vp_\L(\bk)\right)+ O\left(\frac{k^2}{k_{\rm nl}^2} \tilde\eta^{[5]}_{X,\L} X_\L(\bk), \frac{k^2}{k_{\rm nl}^2}  Y^{(3)}_\L(\bk)\right)
     \label{eq:ct5}
\end{align}
where we have included the higher derivative operators involving the quadratic operators $X=\{\vp^2,\,\vp_\gamma,\,\vp_\beta,\,\vp_{\tilde\gamma}\}$, discussed in Sect.~\ref{sub:highder},  and $c^{[5]}_\L= c^{(3)}_\L+c^{(5)}_\L$, $\epsilon^{[5]}_\L\equiv \sum_{n=2}^5\epsilon^{(n)}_\L$,  $c^{[5]}_{X,\L}=c^{(4)}_{X,\L}$  and $\eta^{[5]}_\L\equiv \sum_{n=3}^5\eta^{(n)}_\L$\,. In the last term, $\tilde\eta^{[5]}_{X,\L}$ indicate another stochastic coefficients, while $Y^{(3)}_\L$ denote cubic operators built from $\vp_\L$. 

\subsection{Estimators of the parameters}
\label{sub:estim}
We now  consider the Maximum a Posteriori (MAP) values of $c_\Lambda^{[N]}$ and $\varepsilon_\gamma$, that is the values that maximize the likelihood \eqref{eq:loglikF}. 
Setting to zero the first derivatives with respect to these parameters, we get, at order $N=3$, 
\begin{align}
\bar c_\Lambda^{[3]}&=-\frac{\sum_{\bf n}^{n_{\rm max}}{\rm Re}\left[\left(\delta(\bk_n)-\delta^{[3],{\rm PT}}_{\Lambda}(\bk)\right)\vp_\Lambda(-\bk_n)\right] }{\sum_{\bf n}^{n_{\rm max}} \frac{k_{\bf n}^2}{k_{\rm nl}^2}\left|\vp_\Lambda(\bk_n)\right|^2 }\,,\label{eq:est3-1}\\
\bar \varepsilon_{\gamma,\Lambda}^{[3]}&= \frac{2}{a_\gamma^{(2)}}\,\frac{\sum_{\bf n}^{n_{\rm max}} {\rm Re}\left[\left(\delta(\bk_n)-\delta^{[3],{\rm PT}}_{\Lambda}(\bk)  +\bar c^{[3]}_\L \frac{k_n^2}{k_{\rm nl}^2}\vp_\L(\bk_n) \right) \vp_{\gamma,\Lambda}^{(2)}(-\bk_n)\right]}{\sum_{\bf n}^{n_{\rm max}}\left|\vp_{\gamma,\Lambda}^{(2)}(\bk_n)\right|^2}\,,
\label{eq:est3}
\end{align}
where in both equations we neglected the stochastic terms in Eq.~\eqref{eq:ct3}, under the assumption that they are uncorrelated to $\vp_\L$ and $\vp_{\gamma,\Lambda}^{(2)}$.

At order $N=5$, on the other hand, we have
\begin{align}
\bar c_\Lambda^{[5]}&=-\frac{\sum_{\bf n}^{n_{\rm max}}{\rm Re}\left[\left(\delta(\bk_n)-\delta^{[5]}_\Lambda(\bk_n)-{\rm ``higher \;der."}\right)\vp_\Lambda(-\bk_n)\right] }{\sum_{\bf n}^{n_{\rm max}} \frac{k_{\bf n}^2}{k_{\rm nl}^2}\left|\vp_\Lambda(\bk_n)\right|^2 }\,,\label{eq:est5-1}\\
\bar \varepsilon_{\gamma,\Lambda}^{[5]}&= \frac{2}{a_\gamma^{(2)}}\,\frac{\sum_{\bf n}^{n_{\rm max}} {\rm Re}\left[\left(\delta(\bk_n)-\delta^{[5]}_\Lambda(\bk_n) +\bar c^{[5]}_\L \frac{k_n^2}{k_{\rm nl}^2}\vp_\L(\bk_n)-{\rm ``higher \;der."}\right) \vp_{\gamma,\Lambda}^{(2)}(-\bk_n)\right]}{\sum_{\bf n}^{n_{\rm max}}\left|\vp_{\gamma,\Lambda}^{(2)}(\bk_n)\right|^2}\,,
\label{eq:est5}
\end{align}
where by  ``higher derivative'' terms, we indicate non stochastic terms of type  $O(k_n^2/k_{\rm nl}^2 X_\L)$  and $O(k_n^2/k_{\rm nl}^2 Y^{(3)}_\L)$ in Eq.~\eqref{eq:ct5}. The latter, however, do not correlate with $\vp_{\gamma,\Lambda}^{(2)}$, therefore they do not contribute to Eq.~\eqref{eq:est5} and we will neglect them.

As a first check that our model is properly renormalized, we extract these parameters from grids with different $\Lambda$, and compare the running of the extracted parameters with the ones predicted in PT. 
The simulation measurements and PT predictions are shown respectively in Fig.~\ref{fig:Runcs} as solid and dashed lines, and  the results for $\bar c_\Lambda^{[3]}$ and $\bar c_\Lambda^{[5]}$ are depicted as magenta and blue. 
To ease the comparison, we shifted the the PT curves in such a way as to match the N-body ones at $\L=\L_{\rm uv}=1.07 \;\hMpc$. The difference between the N-body measurement and the (not shifted) PT computation is represented by the dot-dashed lines. Their $\L$-independent values approximate the $\Lambda$-independent quantities $c^{[3]}_{\Lambda_{\rm uv}}$ and $c^{[5]}_{\Lambda_{\rm uv}}$. 

In \cite{Baldauf:2015aha} the speed of sound of the EFTofLSS was estimated from measurements of the nonlinear PS from N-body simulations, as
\beq
\hat c_s^2=-\lim_{k\to 0 }\frac{P_{\rm nl}(k)-P(k)-P_{\rm 1loop}(k)}{2 k^2 P(k) }\,,
\label{eq:cstob}
\eeq
with $P(k)$ is the linear PS and $P_{\rm 1loop}(k)$
the 1-loop contribution. The result at $z=1$ for a WMAP7 cosmology ($\Omega_m=0.272$, $\Omega_\L=0.728$, $n_s=0.967$, $\sigma_8=0.81$) is 
$\hat c_s^2\simeq 0.4 \,{\rm (h^{-1}\,Mpc)}^2$. Since the subtraction is made at 1-loop order, this result should be compared to our $c^{[3]}_{\L_{\rm uv}}$, which, from Fig.~\ref{fig:Runcs}, corresponds to $c^{[3]}_{\L_{\rm uv}}/k_{\rm nl}^2\simeq 0.5 \,{\rm (h^{-1}\,Mpc)}^2$. The two estimates are in reasonable agreement, given the different estimators, cosmologies, and simulation parameters. In particular, notice that the estimator in Eq.~\eqref{eq:cstob} requires an extrapolation to the $k\to 0$ limit, whereas the MAP value, Eq.~\eqref{eq:est3-1}, involves modes up to $k_{\rm max}$.

\begin{figure}
    \centering
    \includegraphics[width=\textwidth]{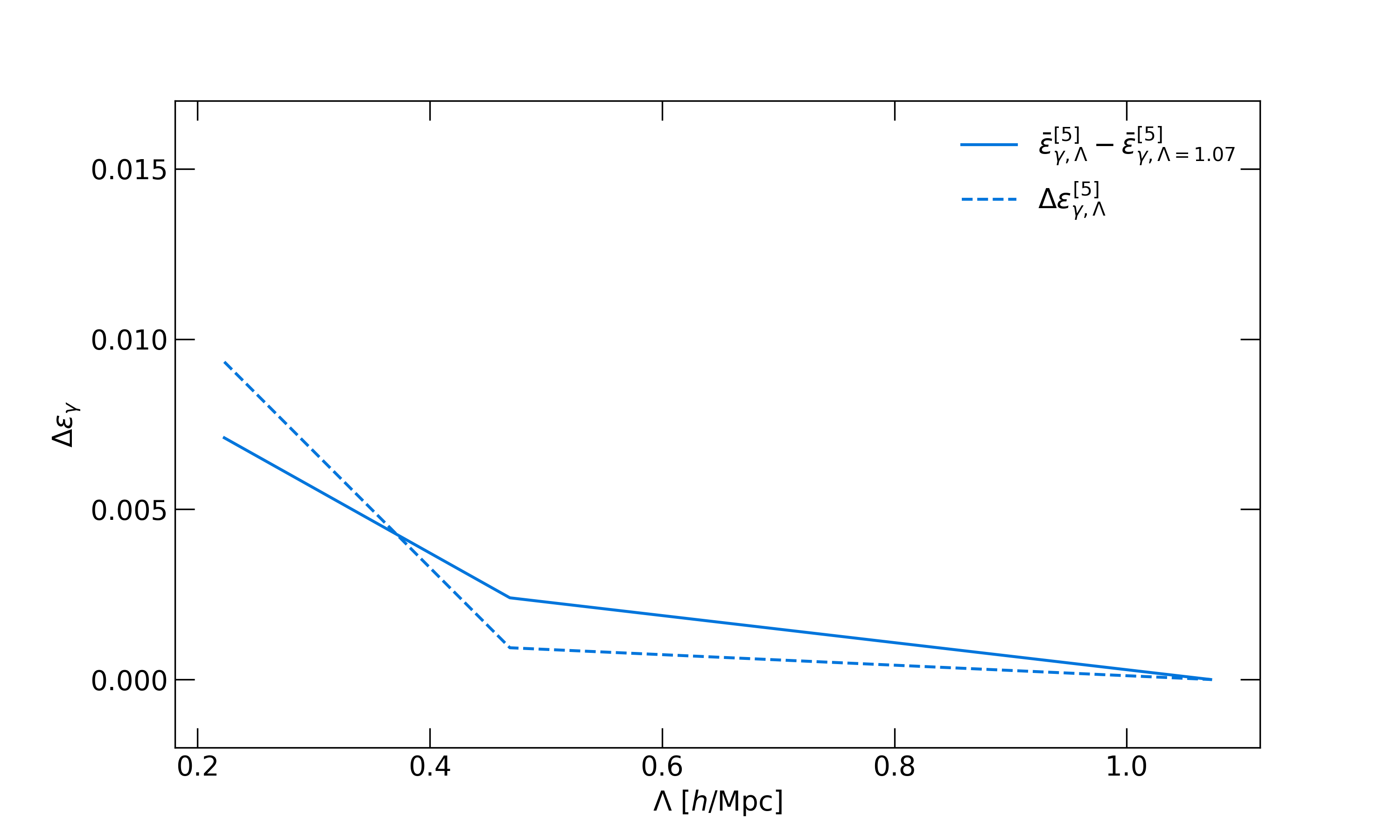}
    \caption{Running of $\varepsilon_\gamma$ at $N=5$ as measured from $N$-body simulations (solid lines) as the difference between the MAP values obtained for each $\Lambda$ from eq. \eqref{eq:est5} and the MAP value at $\Lambda=\L_{\rm uv}=1.07\;\hMpc$. The running predicted by PT, Eq.~\eqref{eq:runcXX}, through Eq.~\eqref{eq:deltaeps}   is shown by the dashed line.}
    \label{fig:Runepsg}
\end{figure}

In computing $\bar c^{[5]}_\L$ we have neglected the higher derivative terms. The agreement between its values and the PT results indicates that they give  sub-leading contributions to the running of this parameter and therefore they could be omitted, along with the stochastic terms, from our models. As we will see, however, this is not  the case for the small parameter $\bar \varepsilon_{\gamma,\Lambda}^{[5]}$. Given its smallness and to the fact that the contributions from higher derivative terms are additive and not multiplicative, it turns out that including these terms improves the fit and removes a residual $\L$-dependence. Therefore, we decide to compute these effects analytically and sum them to the value of 
$\bar \varepsilon_{\gamma,\Lambda}^{[5]}$ extracted from the model without them. It will be done by approximating the sums in \eqref{eq:est5} with integrals and the product of fields with their ensemble averages, 
\begin{align}
\Delta \varepsilon_{\gamma,\L}^{[5], X}&= -\frac{2}{a_\gamma^{(2)}} \frac{c_{X;\Lambda,\Lambda_{\rm uv}}^{[5]}}{k_{\rm nl}^2} \,\frac{\sum_{\bf n}^{n_{\rm max}} k_n^2 \,{\rm Re}\left[X(\bk_n) \vp_\gamma^{(2)}(-\bk_n)\right]}{\sum_{\bf n}^{n_{\rm max}}\left|\vp_\gamma^{(2)}(\bk_n)\right|^2}\,,\nonumber\\
&\to 
-\frac{2}{a_\gamma^{(2)}} \frac{c_{X;\Lambda,\Lambda_{\rm uv}}^{[5]}}{k_{\rm nl}^2} \,\frac{\int^{k_{\rm max}} \frac{d^3 k}{(2\pi)^3} k^2P_{X\gamma}(k)}{\int^{k_{\rm max}} \frac{d^3 k}{(2\pi)^3} P_{\gamma\gamma}(k)}\,,
\label{eq:deltaeps}
\end{align}
where the coefficients $c_{X;\Lambda,\Lambda_{\rm uv}}^{[5]}$ are given in Eq.~\eqref{eq:runningchd}, and we have defined the cross-power spectra,
\beq
P_{X\gamma}(k)\equiv \langle X(\bk) \vp_\gamma^{(2)}(-\bk) \rangle'\,,
\eeq
with  $X=\{\vp^2,\,\vp_\gamma,\,\vp_\beta,\,\vp_{\tilde\gamma}\}$, the operators appearing in the higher derivative terms, Eq.~\eqref{eq:highder}. 
Notice that we will consider only the contributions to $\Delta \varepsilon_{\gamma,\L}^{[5], X}$ coming from the perturbative running from $\L_{\rm uv}$ to $\L$, which is equivalent to setting $c^{[5]}_{X,\L_{\rm uv}}=0$. In Fig.~\ref{fig:Runepsg} we compare the $\L$ dependence of the $\varepsilon^{[5]}_{\gamma,\L}$ parameter as obtained from the MAP estimator \eqref{eq:est5} from N-body simulations (solid line) with the one obtained from Eq.~\eqref{eq:deltaeps} assuming the PT running of Eq.~\eqref{eq:runcXX} (dashed line).

The largest wavenumber, $k_{\rm max}$, included in the likelihood is a  crucial parameter. Intuitively, it must be fixed at a value well inside the range of validity of the PT order we are considering. More specifically, since we are interested in extracting the bootstrap parameter $\varepsilon_\gamma$, we notice that the RHS of Eq.~\eqref{eq:est3} contains the ratio of a 2-loop ($O(P^3)$) to a 1-loop  ($O(P^2)$) term,  that is, it is formally of linear order ($O(P)$), whereas, by the same counting,  Eq.~\eqref{eq:est5}  is formally a 1-loop quantity. Therefore, we anticipate that the appropriate $k_{\rm max}$ should be comparable to that of linear PT for $\bar \varepsilon_\gamma^{[3]}$ and to that of 1-loop PT for $\bar \varepsilon_\gamma^{[5]}$. In the next section, we will discuss this issue more quantitatively.

\begin{figure}
    \centering
    \includegraphics[width=\textwidth]{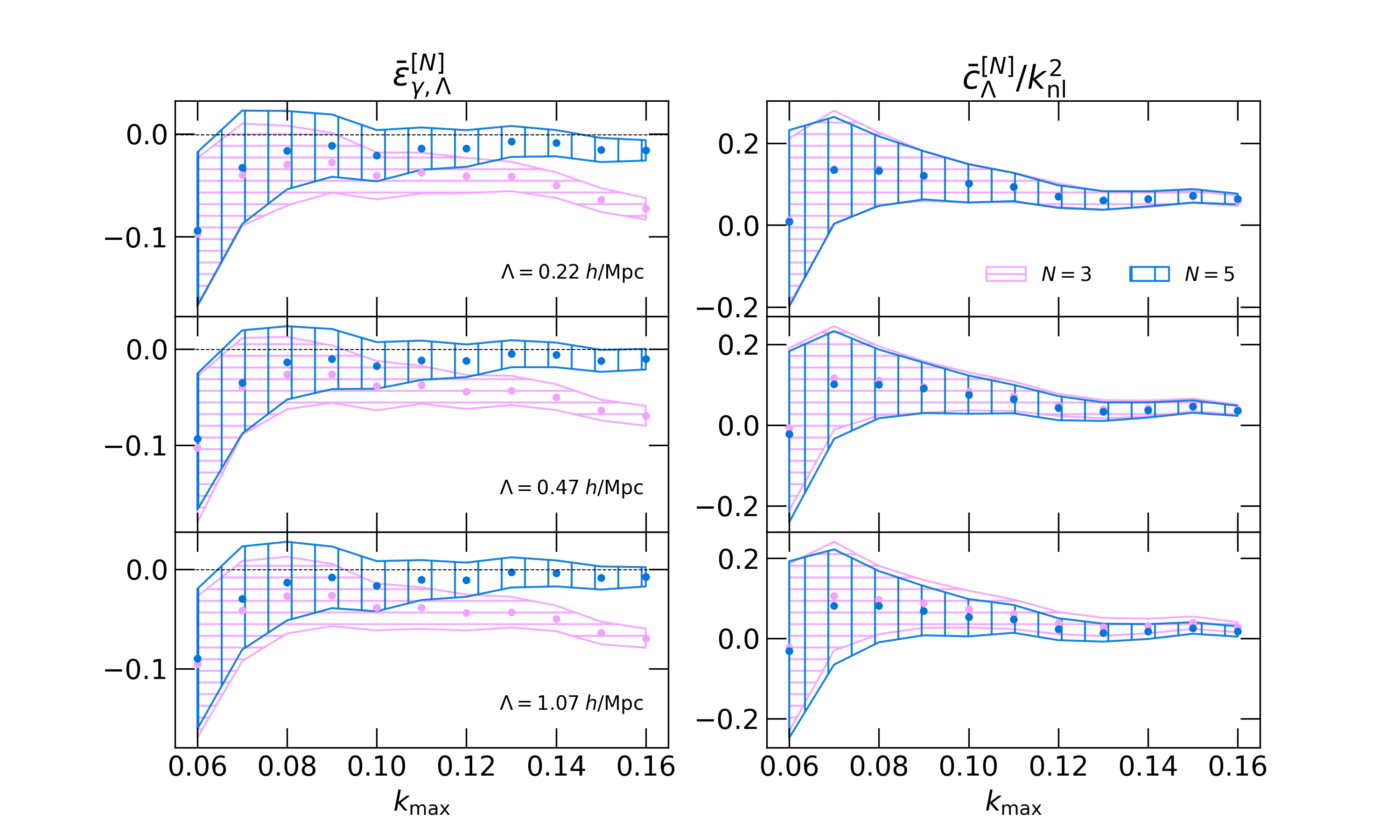}
    \caption{$k_{\rm max}$ dependence of the maximum likelihood parameters $\bar\varepsilon^{[N]}_{\gamma,\L}$ and $\bar c^{[N]}_{\L}/k^2_{\rm nl}$ for $N=3$ (pink area) and $N=5$ (blue area). In the $N=5$ case, higher derivatives contributions have not been included. }
    \label{fig:kmax}
\end{figure}
\section{Results}
\label{sect:results}
\begin{figure}
    \centering
\includegraphics[width=0.48\textwidth]{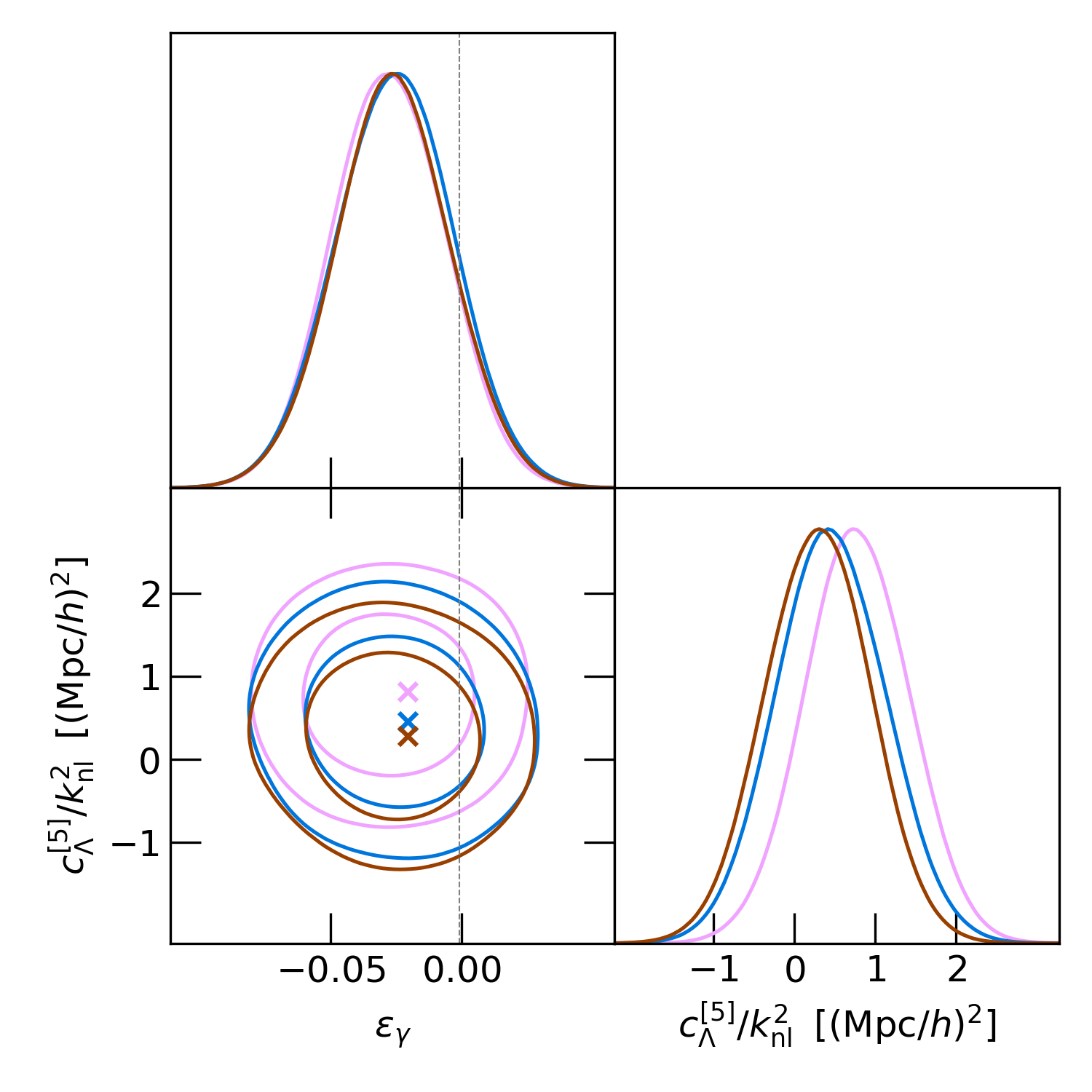}
\includegraphics[width=0.48\textwidth]{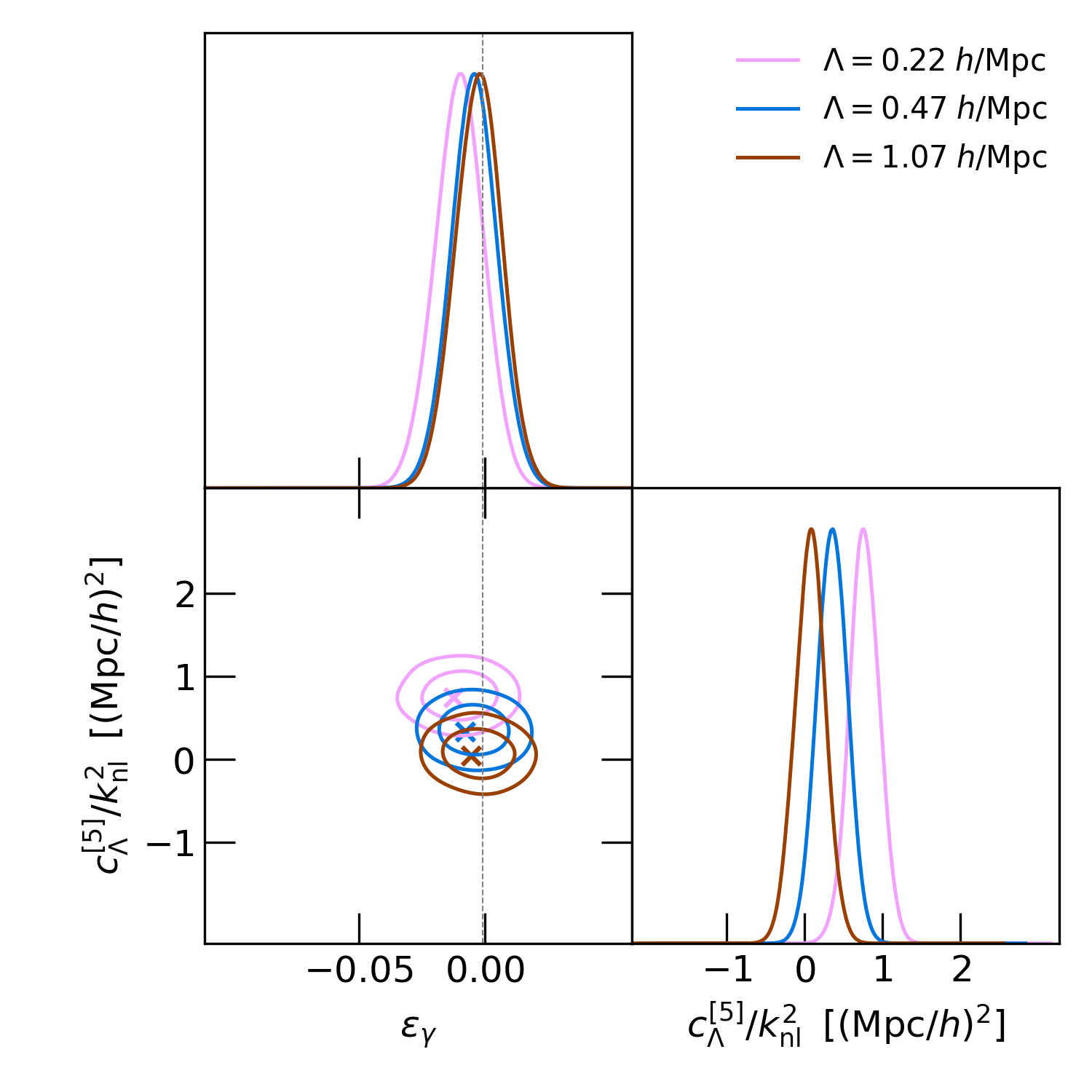}
\includegraphics[width=0.48\textwidth]{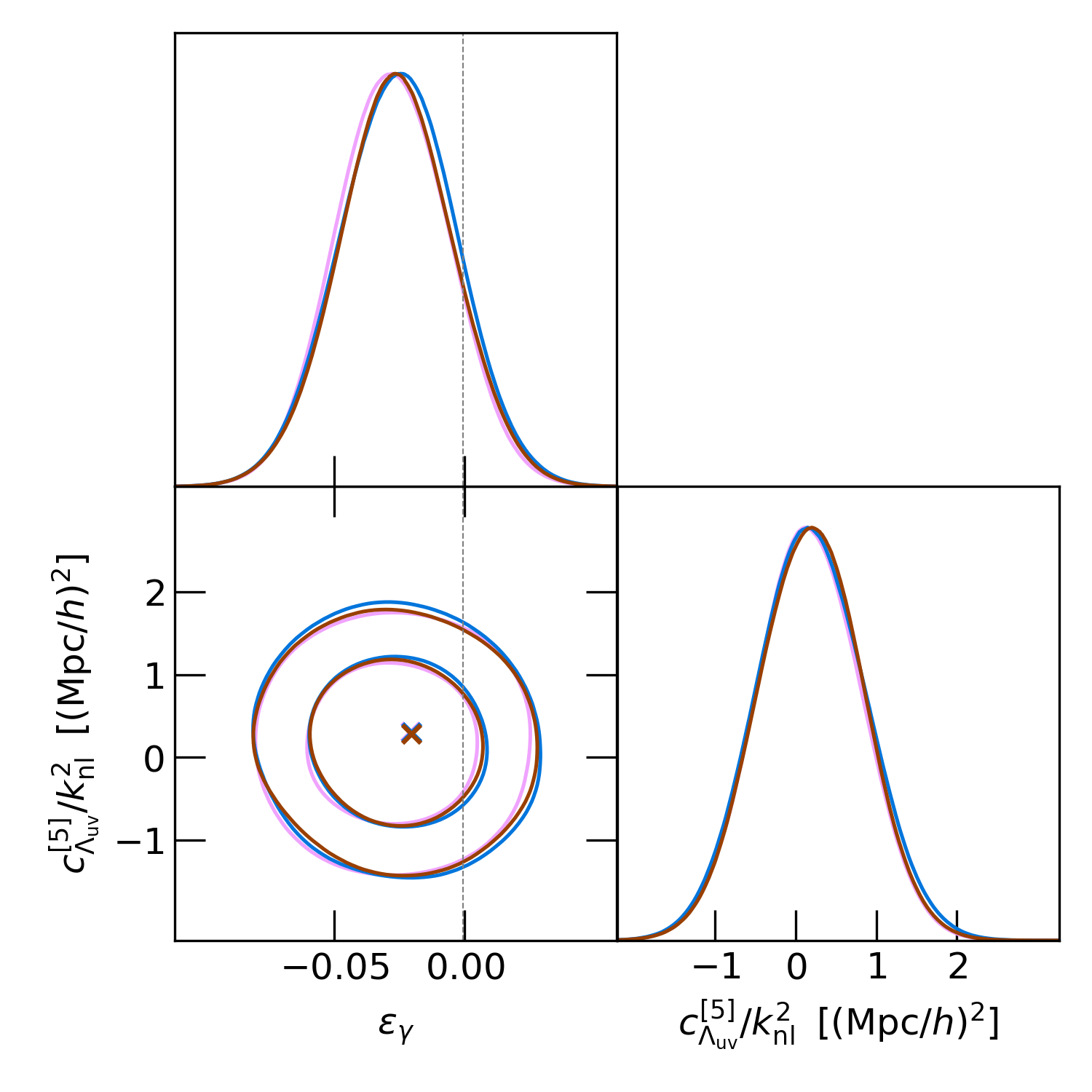}
\includegraphics[width=0.48\textwidth]{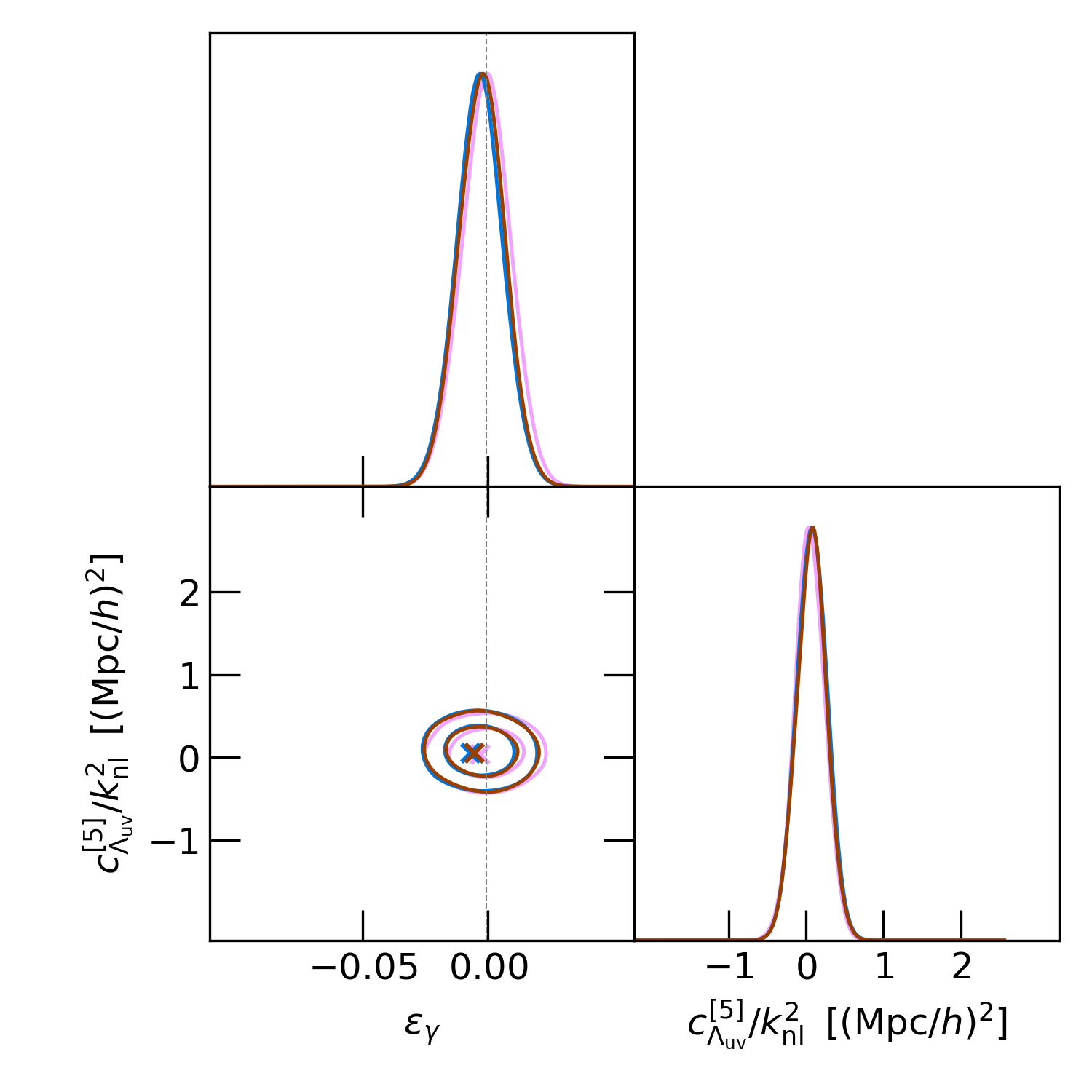}
    \caption{Results in terms of $\Lambda$-dependent (upper row) and  $\Lambda$-independent parameters (lower row). The left (right)  column is obtained at order $N=3$ ($N=5$). The crosses indicate the MAP values found from the estimators in equations \eqref{eq:est3-1}--\eqref{eq:est5} using the N-body data without added noise. The $k_{\rm max}$ values used in these plots are $0.09\;\hMpc$ for $N=3$ and $0.14\;\hMpc$ for $N=5$. Notice that, in order to compare the two models, we plot contours in the $\epsilon_\gamma -c^{[5]}_\Lambda/k_{\rm nl}^2$ plane also in the $N=3$ case,  using Eq.~\eqref{eq:c3c5uv} to relate $c^{[3]}_\Lambda/k_{\rm nl}^2$ to $c^{[5]}_\Lambda/k_{\rm nl}^2$. }
    \label{fig:triangle}
\end{figure}

\begin{table}
    \centering
    \label{tab:Lambda_vs_gridsize}
    \resizebox{0.5\textwidth}{!}{
    \begin{tabular}{|c||c|c|c|}
        \hline
        $\Lambda\;[\hMpc]$     & $0.223$ & $0.469$ & $1.072$ \\ \hline
        $N_g$ ($N=3$) & $141$    & $298$    & $682$    \\ \hline
        $N_g$ ($N=5$) & $212$    & $447$    & $1024$   \\ \hline
    \end{tabular}
    }
    \caption{Correspondence between $\Lambda$ and $N_g$}
\end{table}

Summarizing the discussion in Sect.~\ref{sect:Likelihood} we  consider two renormalized models: $N=3$ and $N=5$ where, in both cases, we will use 
        \begin{equation}
            \delta^{[N]}(\bk)= \delta^{[N],{\rm PT}}_\L(\bk)+ \frac{1}{2} \varepsilon_\gamma \,a^{(2)}_{\rm EdS} \,\vp^{(2)}_{\gamma,\Lambda}(\bx)-c^{[N]}_\L \frac{k^2}{k_{\rm nl}^2}\vp_\L(\bk)\,,
            \label{eq:finalmod}
        \end{equation}
and, in the case of $N=5$, we will add the higher derivative contributions \eqref{eq:deltaeps}, computed analytically, to the value of $\varepsilon_\gamma$ inferred from the data.  We  also checked the irrelevance of stochastic contributions of Eqs.~\eqref{eq:ct3} and \eqref{eq:ct5}.  This is explained by the observation that stochastic terms do not contribute to \eqref{eq:est3} and \eqref{eq:est5}.

Now we describe the results obtained from the full likelihood, Eq.~\eqref{eq:loglikF}. We consider three different values for the cutoff on the linear field: $\Lambda/(\hMpc)=0.22,\,0.47,\,1.07$ (see Table~\ref{tab:Lambda_vs_gridsize}). The largest one, $1.07$, will effectively be our $\Lambda_{\rm uv}$.

We consider the dependence of the extracted $\varepsilon_\gamma$ on varying $k_{\rm max}$
%\MP{(TO BE ADDED WITH SHOT NOISE AVERAGING) for one of the noise realizations}
, and stop when its deviation from the true value is equal to 1$\sigma$.
This procedure is illustrated in the left panels of Fig.~\ref{fig:kmax}, for $N=3$ (blue areas) and $N=5$ (orange areas), for three different values of the linear cutoff $\L$. This criterion gives $k_{\rm max}^{N=3}=0.09 \,\hMpc$, $k_{\rm max}^{N=5}=0.14 \,\hMpc$, respectively, for $\Lambda \agt 0.4 \, \hMpc$. The right panels of Fig.~\ref{fig:kmax} show the dependence on $k_{\rm max}$ of $c^{[N]}_\L$. 

The inferred values for $\varepsilon_\gamma$ and $c^{[5]}_\Lambda/k_{\rm nl}^2$ for these $k_{\rm max}$ values are given in the upper plots of Fig.~\ref{fig:triangle}, for $N=3$ (left) and $N=5$ (right), without including the higher derivative contributions to $\varepsilon_\gamma$. The posterior distributions shown are averaged over the ten shot noise realizations, that is their maximum is the average of the ten maximums at fixed $N$ and $\Lambda$ values.   Notice that, in order to compare the two models, we plot contours in the $\epsilon_\gamma -c^{[5]}_\Lambda/k_{\rm nl}^2$ plane also in the $N=3$ case,  using Eq.~\eqref{eq:c3c5uv} to relate $c^{[3]}_\Lambda/k_{\rm nl}^2$ to $c^{[5]}_\Lambda/k_{\rm nl}^2$.

In these figures, we also display with crosses color-coded as the posteriors, the MAP values for the parameters, derived from Eqs.~\eqref{eq:est3} and \eqref{eq:est5} using the same datasets and fields, without the inclusion of artificial noise.

The lower plots in Fig.~\ref{fig:triangle} show that our models are properly renormalized. We express the $\L$-dependent parameter $c^{[5]}_\Lambda/k_{\rm nl}^2$ for  $\Lambda=0.22 \;\hMpc$ and $\Lambda=0.47 \;\hMpc$, in terms of the $\Lambda$-independent one, $c^{[5]}_{\Lambda_{\rm uv}}/k_{\rm nl}^2$, assuming the perturbative running of Eqs.~\eqref{eq:runc3} and \eqref{eq:runningc5}. Moreover, we include the higher derivative contribution to $\varepsilon_\gamma$,  using Eq.~\eqref{eq:deltaeps}.

The contours and the MAP parameters are now much better aligned for all three $\L$ values. In particular, the role of higher derivative corrections in reducing the $\L$ dependence can be quantified comparing the plots at the first and second lines in the $\varepsilon_\gamma$ direction.

Overall, the precision on the parameter $\varepsilon_\gamma$ increases by a factor $2.4$ for the $N=5$ model with respect to the $N=3$ one. This improvement is due to the larger $k_{\rm max}$ attainable at higher perturbative orders.

\section{Discussion and conclusions}
\label{Sect:discussion}

In this work, we have taken a first step toward implementing field-level inference beyond the standard $\Lambda$CDM model, with a particular focus on optimizing precision tests in the nonlinear regime of LSS. As a concrete example of how model-independent constraints on new physics can be derived, we focused on the bootstrap coefficient of the second-order PT kernel for matter in real space. 

A central aspect of our analysis is the detailed treatment of the UV cutoff dependence inevitably induced by discretizing the field on a grid. Field theories on a lattice are well studied in other areas of physics, where the challenge of defining a meaningful continuum limit is well recognized and established methods to address it have been developed. Discretization naturally introduces a UV cutoff $\Lambda$, associated either with the removal of spurious aliasing effects or, ultimately, with the Nyquist frequency. The field theory must therefore be properly renormalized to ensure that physical predictions are independent of grid-induced artifacts.

Throughout this work, we restricted our analysis to the perturbative regime, considering only modes with wavenumbers $k < k_{{\rm max}} \ll k_{{\rm nl}}$, where $k_{{\rm nl}}$ denotes the nonlinear scale. The UV theory is defined at a cutoff scale $\Lambda_{{\rm uv}} \gg k_{{\rm nl}}$, ensuring that higher-derivative corrections are controlled by a single physical scale, $k_{{\rm nl}}$, while $O(k^2/\L_{\rm uv}^2)$ terms are ``irrelevant'', since the values of their coefficients affect the models at lower cutoff $\L$ only via contributions suppressed by powers of $\L^2/\L^2_{\rm uv}$. The evolution from the UV down to lower cutoff scales $\Lambda < \Lambda_{{\rm uv}}$ can then be computed perturbatively, and therefore it does not necessarily rely on the validity of a derivative expansion in powers of $k^2/\Lambda^2$. At any given $\Lambda$, the model is expressed in terms of cutoff fields $\varphi_\Lambda$, and consists of a truncated PT series supplemented by a set of counterterms. The impact of modes with $k > \Lambda$ is encoded in the counterterm coefficients, while modes with $k < \Lambda$ are treated using \texttt{GridSPT}, which we extended to implement the bootstrap parameterization. Lowering the value of $\L$ in a controlled way is instrumental in reducing the number of grid points, and consequently, in reducing the computational cost.

We studied two renormalized models: one including terms up to $N = 3$ and another up to $N = 5$, where $N$ denotes the perturbative order. For both cases, we derived the corresponding $k_{{\rm max}}$ values, finding them to be consistent with those obtained in standard EFTofLSS analyses for correlators \cite{Nishimichi:2020tvu}. In the $N=5$ case, we demonstrated that including the contributions from higher-derivative terms quadratic in the linear fields is essential for achieving $\Lambda$-independent results and for obtaining an unbiased extraction of the bootstrap coefficient. 

In~\cite{Assassi:2014fva} a non-renormalization theorem for the coefficients of galileon operators was proven. As $\vp_\gamma$ belongs to this category, requiring renormalization might seem, at first sight, in contraddiction with this theorem. Nevertheless, the theorem is inapplicable to the coefficients of higher derivative terms like $(\partial^2/k_{\rm nl}^2 )\vp_{\gamma,\L}$, such as those examined in this study, therefore there is no contraddiction.  

The methodology developed here is readily extendable to more realistic scenarios. Incorporating biased tracers is conceptually and practically straightforward within this framework, requiring only the addition of further operators to the \texttt{GridSPT} setup, similarly to what we implemented here with the bootstrap operator. Bias operators are in general not protected by momentum conservation, so the renormalization of bias parameters starts at $O(k^0)$ and not at $O(k^2)$ as for the bootstrap parameters \cite{Assassi:2014fva, Rubira:2023vzw, Nikolis:2024kbx}. Furthermore, redshift-space distortions, already treated perturbatively within the \texttt{GridSPT} framework in previous work \cite{Taruya2022} (for other field-level approaches at field level in redshift space see also \cite{Obuljen:2022cjo, Stadler:2024aff}), can be included, although modeling Fingers-of-God (FoG) effects accurately remains a challenging aspect, as discussed extensively in the literature (see, e.g., \cite{Scoccimarro:2004tg,Taruya:2010mx, Eggemeier:2025xwi}).
Finally, the application to real data will require a full  sampling over the cosmological parameters and marginalization over the amplitudes and phases of the linear fields. As shown in \cite{Nguyen2024} this task can be accomplished via Hamiltonian Monte Carlo and other advanced sampling techniques \cite{Neal:1993, Neal:2003, neal2011mcmc}.

\section*{Acknowledgements}
\noindent
We thank  G.~Biselli, P.~Conzinu, G.~D'Amico, M.~Garny, D.~Jeong, D.~Linde,  K.~Osato, K.~Pardede, H.~Rubira, and F.~Schmidt for useful discussions.
M. Pietroni acknowledges support by the MIUR Progetti di Ricerca di Rilevante Interesse Nazionale (PRIN) Bando 2022 - grant 20228RMX4A, funded by the European Union - Next generation EU, Mission 4, Component 1, CUP C53D23000940006. This work was supported in part by MEXT/JSPS KAKENHI Grant Numbers 
JP20H05861 and JP23K20844
(AT and TN), JP23K25868 (AT), JP22K03634,
JP24H00215, and JP24H00221 (TN). This research was also supported by the Munich Institute for Astro-, Particle and BioPhysics (MIAPbP) which is funded by the Deutsche Forschungsgemeinschaft (DFG, German Research Foundation) under Germany´s Excellence Strategy – EXC-2094 – 390783311. Numerical computations were carried out on Cray XC50 at Center for Computational Astrophysics, National Astronomical Observatory of Japan, and on the High Performance Computing facility of the University of Parma, Italy.

\appendix

\section{Second order bootstrap effect at third order}
\label{app:third}

As briefly introduced in section \ref{sect:GridSPT}, the effect of $a_\gamma^{(2)}$ at third order has been found to be negligible, so the results presented in this paper derive from an implementation of bootstrap only at second order. This appendix supplements the main results by providing more details about implementation and numerical proof of the marginal contributions of third order to the estimation of $\varepsilon_\gamma$.

Continuing on the same line as equations \eqref{eq:delta2}--\eqref{eq:phigamma}, at third order in Fourier space we have

\begin{align}
\delta^{(3)}(\bk,a) &= {\cal I}_{\bk;\bq_1,\bq_2,\bq_3} F_{3}(\bq_1,\bq_2,\bq_3; a) \varphi(\bq_1, a)\varphi(\bq_2, a)\varphi(\bq_3, a)\,,\\
\theta^{(3)}(\bk,a) &= {\cal I}_{\bk;\bq_1,\bq_2,\bq_3} G_{3}(\bq_1,\bq_2,\bq_3; a) \varphi(\bq_1, a)\varphi(\bq_2, a)\varphi(\bq_3, a)\,,
\end{align}

which leads to the definition of the following fields: 

\begin{align}
\varphi^{(3)}_{\beta\beta}(\bx)\equiv &\,\frac{1}{2}\left[
\partial_i \varphi(\bx) \left(\frac{\partial_i}{\partial^2}\varphi^{(2)}_\beta(\bx)\right) + \partial_i \varphi^{(2)}_\beta(\bx) \left(\frac{\partial_i}{\partial^2}\varphi(\bx)\right)
\right]\\
&+\left(\frac{\partial_i\partial_j}{\partial^2}\varphi(\bx)\right)\left(\frac{\partial_i\partial_j}{\partial^2}\varphi^{(2)}_\beta(\bx)\right)\,,\nonumber\\
\varphi^{(3)}_{\beta\gamma}(\bx)\equiv &\,\frac{1}{2}\left[
\partial_i \varphi(\bx) \left(\frac{\partial_i}{\partial^2}\varphi^{(2)}_\gamma(\bx)\right) + \partial_i \varphi^{(2)}_\gamma(\bx) \left(\frac{\partial_i}{\partial^2}\varphi(\bx)\right)
\right]\\
&+\left(\frac{\partial_i\partial_j}{\partial^2}\varphi(\bx)\right)\left(\frac{\partial_i\partial_j}{\partial^2}\varphi^{(2)}_\gamma(\bx)\right)\,,\nonumber
\\ \varphi^{(3)}_{\gamma\beta}(\bx)\equiv &\,\varphi(\bx)\varphi^{(2)}_\beta(\bx)- \left(\frac{\partial_i\partial_j}{\partial^2}\varphi(\bx)\right)\left(\frac{\partial_i\partial_j}{\partial^2}\varphi^{(2)}_\beta(\bx)\right)\,,\\
\varphi^{(3)}_{\gamma\gamma}(\bx)\equiv &\,\varphi(\bx)\varphi^{(2)}_\gamma(\bx)- \left(\frac{\partial_i\partial_j}{\partial^2}\varphi(\bx)\right)\left(\frac{\partial_i\partial_j}{\partial^2}\varphi^{(2)}_\gamma(\bx)\right)\,,\\
\varphi^{(3)}_{\alpha_a\gamma}(\bx)\equiv & \,\left(\frac{\partial_i}{\partial^2}\varphi^{(2)}_\gamma(\bx)\right)\partial_i \varphi(\bx)- \partial_i \varphi^{(2)}_\gamma(\bx) \left(\frac{\partial_i}{\partial^2}\varphi(\bx)\right)\,.
\end{align}

Finally we find:

\begin{align}
\label{delta3}
   \delta_3(\bx) & =\varphi^{(3)}_{\beta\beta}(\bx)+
\bigg(\frac{1}{2}a_{\gamma a,{\rm EdS}}^{(3)} -a_{\gamma b,{\rm EdS}}^{(3)} + 1\bigg)\varphi^{(3)}_{\gamma\beta}(\bx) +\left(\frac{1}{8} a_{\gamma a,{\rm EdS}}^{(3)}-\frac{1}{4}a_{\gamma b,{\rm EdS}}^{(3)} \right)\varphi^{(3)}_{\alpha_a\gamma}(\bx)\nonumber\\
 & +\left(\frac{1}{2}a_{\gamma b,{\rm EdS}}^{(3)}-\frac{1}{4}a_{\gamma a,{\rm EdS}}^{(3)}+a_{\gamma}^{(2)} - 1\right) \varphi^{(3)}_{\beta\gamma}(\bx) +\left(\frac{1}{4}a_{\gamma a,{\rm EdS}}^{(3)}+\frac{1}{2}a_{\gamma b,{\rm EdS}}^{(3)}\right)\varphi^{(3)}_{\gamma\gamma}(\bx)\,,\\
   \theta_3(\bx) & =\varphi^{(3)}_{\beta\beta}(\bx)+
\bigg(\frac{1}{2}d_{\gamma a,{\rm EdS}}^{(3)} -d_{\gamma b,{\rm EdS}}^{(3)} + 1\bigg)\varphi^{(3)}_{\gamma\beta}(\bx) +\left(\frac{1}{8} d_{\gamma a,{\rm EdS}}^{(3)}-\frac{1}{4}d_{\gamma b,{\rm EdS}}^{(3)} \right)\varphi^{(3)}_{\alpha_a\gamma}(\bx)\nonumber\\
 & +\left(\frac{1}{2}d_{\gamma b,{\rm EdS}}^{(3)}-\frac{1}{4}d_{\gamma a,{\rm EdS}}^{(3)}+d_{\gamma}^{(2)} - 1\right) \varphi^{(3)}_{\beta\gamma}(\bx) +\left(\frac{1}{4}d_{\gamma a,{\rm EdS}}^{(3)}+\frac{1}{2}d_{\gamma b,{\rm EdS}}^{(3)}\right)\varphi^{(3)}_{\gamma\gamma}(\bx)\,.
\end{align}

Setting all the third order bootstrap parameters to their ${\rm EdS}$ value is equivalent to modifying the third order of \texttt{GridSPT} in the following way:

\begin{equation}
    \delta^{(3)}(\bx)=\delta^{(3)}_{\rm EdS}(\bx)+\varepsilon_\gamma a_{\gamma,{\rm EdS}}^{(2)}\left(\varphi_{\beta\gamma}^{(3)}(\bx)+\frac{1}{2}\varphi_{\gamma\gamma}^{(3)}(\bx)\right)\,.
\end{equation}

Assuming the velocity field to be irrotational, the tensor products appearing in the fields $\varphi_{\beta\gamma}^{(3)}$ and $\varphi_{\gamma\gamma}^{(3)}$ can be simplified to make numerical computation feasible. This leads to the following expressions:

\begin{gather}
    \varphi_{\beta\gamma}^{(3)}(\bx)=\frac{1}{2}\partial^2\left(\frac{\partial_i}{\partial^2}\varphi(\bx)\frac{\partial_i}{\partial}\varphi_\gamma^{(2)}(\bx)\right)\,, \\
     \varphi_{\gamma\gamma}^{(3)}(\bx)=\varphi(\bx)\varphi_\gamma^{(2)}(\bx)-\frac{1}{2}\partial^2\left(\frac{\partial_i}{\partial^2}\varphi(\bx)\frac{\partial_i}{\partial}\varphi_\gamma^{(2)}(\bx)\right)+\frac{1}{2}\partial_i\varphi(\bx)\frac{\partial_i}{\partial}\varphi_\gamma^{(2)}(\bx)+\partial_i\varphi_\gamma^{(2)}(\bx)\frac{\partial_i}{\partial^2}\varphi(\bx)\,.
\end{gather}

\begin{figure}
    \centering
    \includegraphics[width=0.48\textwidth]{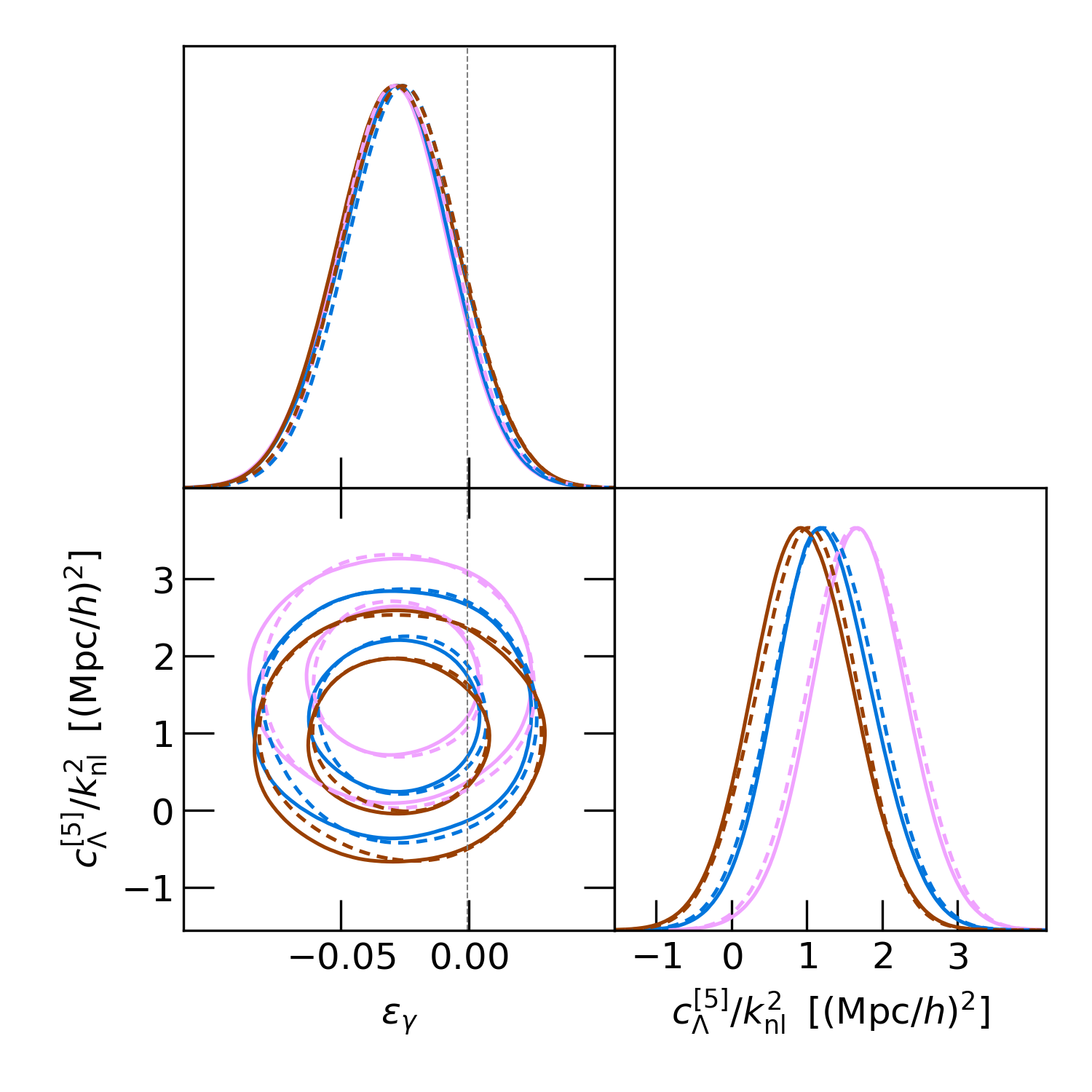}
    \includegraphics[width=0.48\textwidth]{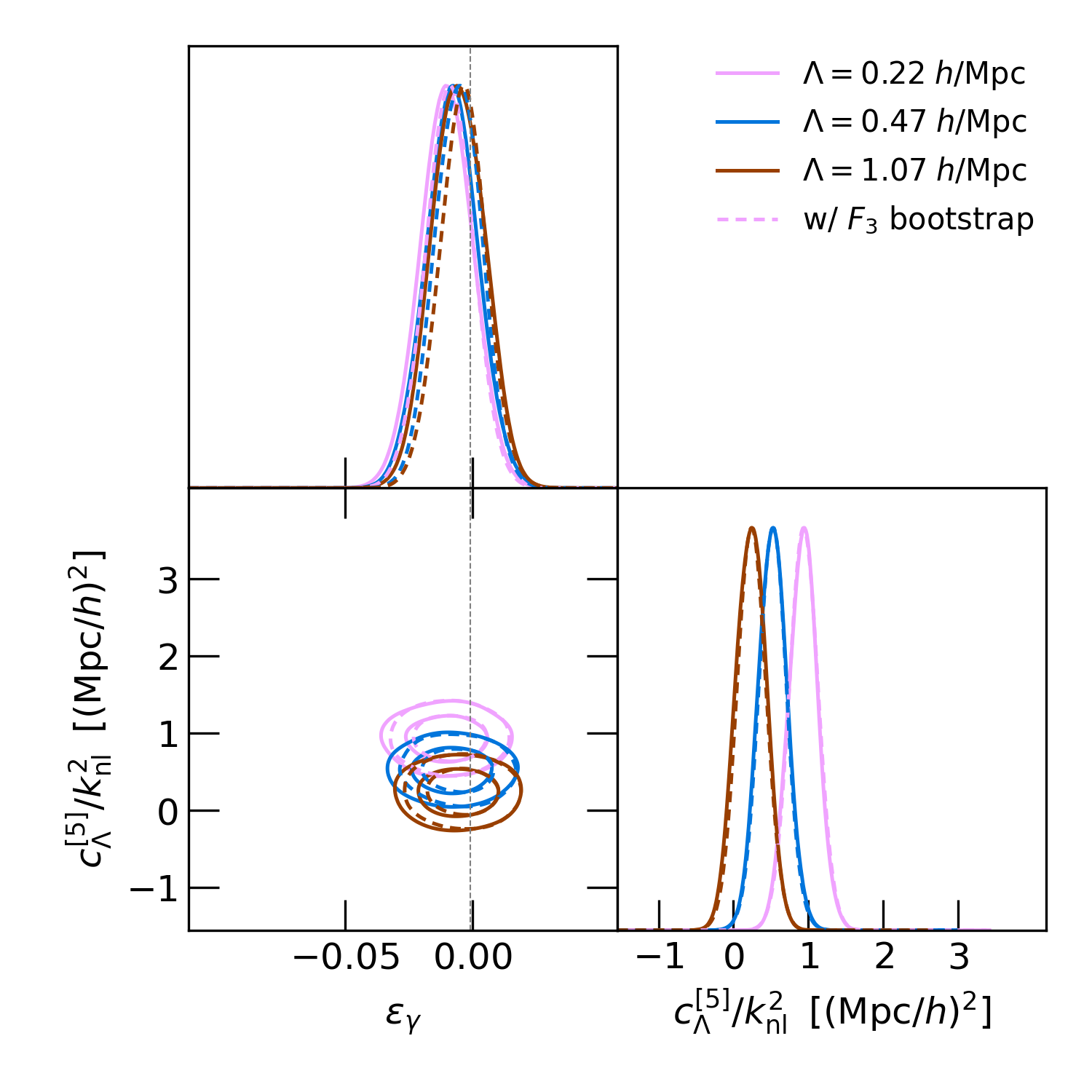}
    \caption{Comparison between adding the second order bootstrap only up to second order (solid lines) or also to third order (dashed lines). The left (right)  column is obtained at order $N=3$ ($N=5$).}
    \label{fig:triangle_F3}
\end{figure}

Figure \ref{fig:triangle_F3} compares the countours obtained when adding the contribution to third order. while the errorbars appear to improve the effect is very marginal, showing how higher orders give minimal contributions to the detection of alternative cosmologies using the bootstrap approach.

\section{On injecting shot noise in a continuous field}
\label{app:noise}

In cosmology, the noise associated with samples of observed galaxies or other cosmological probes follows a Poisson distribution, because galaxies do (excluding small scale hydrodynamical physics and other non-gravitational effects) sample the underlying matter density distribution $\rho(\bx)$.
In principle this type of noise can be mimicked by starting from the number count:

\begin{equation}
\label{eq:counts}
    \tilde{N}(\bx) \equiv \rho_m(\bx)\;\Delta^3x=\left(1+\delta_m(\bx)\right)\bar{n}\;\Delta^3x\,,
\end{equation}

where $\delta_m(\bx)\equiv\frac{\rho_m(\bx)}{\langle\rho_m\rangle}-1$ is the density contrast of dark matter and $\Delta^3x$ is a finite volume element.
This number can be used as the position-dependent average of a Poisson distribution from which to sample an actual number of particles. If the density at $\bx$ is high then more particles are likely to be sampled, and viceversa:

\begin{equation}
    N(\bx) \sim P(N\;|\;\tilde{N})=\frac{\tilde{N}^{N}e^{-\tilde{N}}}{N!}\,;
\end{equation}

note how the distribution of $N$ is conditional on $\tilde{N}$, because the latter is also a random variable. Given $N(\bx)$, the noisy density contrast $\delta(\bx)$ can be obtained by simply rescaling the counts by their mean over every position $\bx$ so that $\langle\delta(\bx)\rangle=0$.
It is known that in galaxies or other tracers, the noise is independent from the signal and additive. To show that this is the case here as well, one just needs to prove the following:

\begin{equation}
\label{eq:shot_noise_proof-1}
    \left\langle\delta_m(\bx)\left[\delta(\bx')-\delta_m(\bx')\right]\right\rangle\stackrel{?}{=}0\,.
\end{equation}

Using the definitions of expected value and the axioms of probability (and omitting the dependence on $\bx$ for the sake of clarity):

\begin{align}
\label{eq:shot_noise_proof-2}
    \left\langle\left(\frac{\rho_m}{\langle\rho_m\rangle}-1\right)\left(\frac{\rho}{\bar{n}}-1\right)\right\rangle & =\left\langle\left(\frac{\rho_m}{\langle\rho_m\rangle}-1\right)\left(\frac{\rho_m}{\langle\rho_m\rangle}-1\right)\right\rangle \\ \nonumber
    \frac{1}{\langle\rho_m\rangle\bar{n}}\frac{1}{V^2}\langle \tilde{N}\;N\rangle-1 & = \frac{1}{\langle\rho_m\rangle^2}\frac{1}{V^2}\langle \tilde{N}\;\tilde{N}\rangle-1 \\ \nonumber
    \frac{1}{\bar{n}}\iint{\rm D}\tilde{N}\;{\rm D}N\;\tilde{N}NP(\tilde{N},\,N) & = \frac{1}{\langle\rho_m\rangle}\int{\rm D}\tilde{N}\;\tilde{N}^2P(\tilde{N}) \\ \nonumber
    \int{\rm D}\tilde{N}\;\tilde{N}\left[\int{\rm D}N\;NP(N\;|\;\tilde{N})\right]P(\tilde{N}) & = \int{\rm D}\tilde{N}\;\tilde{N}^2P(\tilde{N})\,.
\end{align}

\begin{figure}
    \centering
    \includegraphics[width=\textwidth]{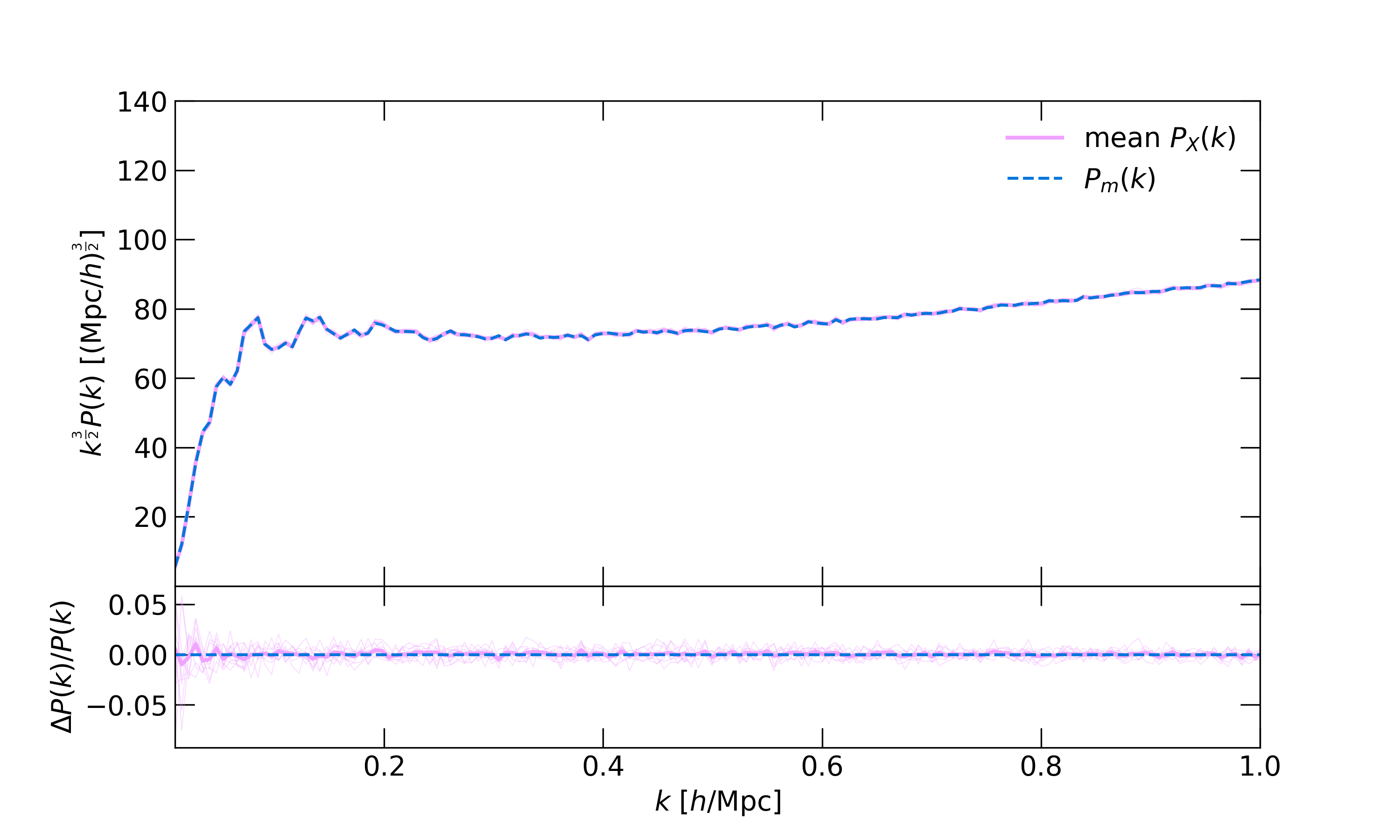}
    \caption{Comparison between the cross power spectrum $P_X(k)\equiv\frac{\delta_D(k-k')}{(2\pi)^3}\langle\delta_m(\bk)\delta(\bk')\rangle$ with and without noise in ten independent realizations of the noise (pink thin lines), and the auto power spectrum of dark matter $P_m(k)$ (blue dashed line). The mean of the ten realization is shown as the pink thick line.}
    \label{fig:shot_noise_proof}
\end{figure}

\begin{figure}
    \centering
    \includegraphics[width=\textwidth]{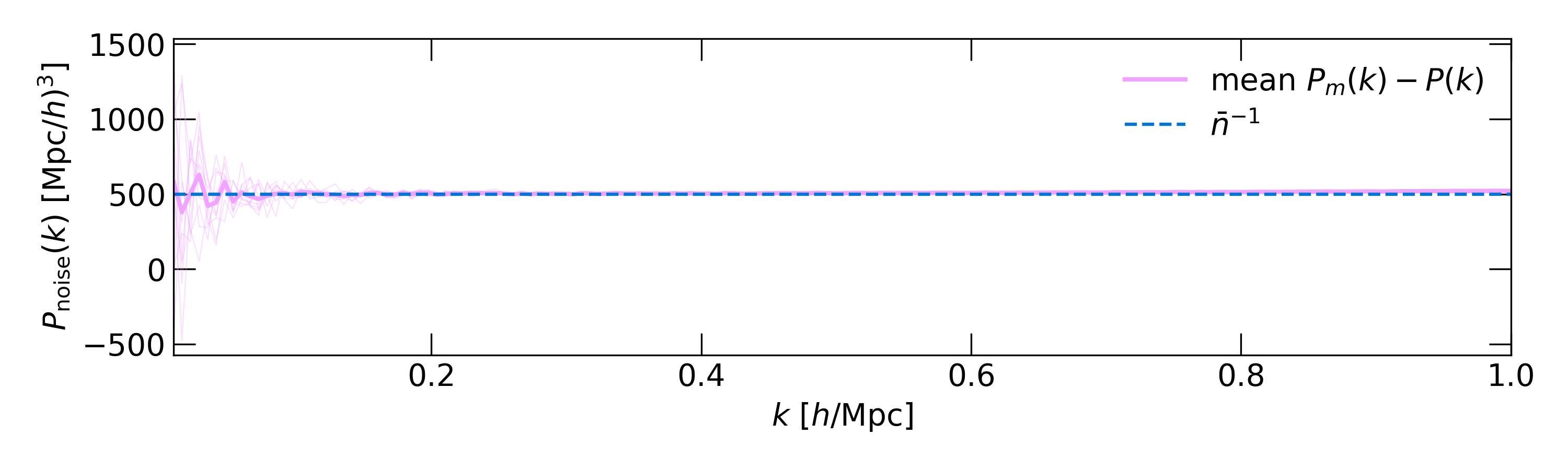}
    \caption{Comparison between the difference between the auto power spectra of dark matter $P_m(k)$ and the noisy counterpart $P(k)$ in ten independent realizations of the noise (pink thin lines), and the theoretical value of noise in case its distribution is Poissonian (blue dashed line). The mean of the ten realization is shown as the pink thick line.}
    \label{fig:shot_noise_theory}
\end{figure}

Notice that $\langle\rho_m\rangle=\bar{n}$ follows naturally from equation \eqref{eq:counts}. To complete the proof, the integral within square brackets must be equal to $\tilde{N}$:

\begin{align}
    \tilde{N} & =\int{\rm D}N\;NP(N\;|\;\tilde{N}) \\ \nonumber
    & = \sum_{N=0}^\infty N\frac{\tilde{N}^{N}e^{-\tilde{N}}}{N!} \\ \nonumber
    & = \tilde{N}e^{-\tilde{N}}\sum_{N=1}^\infty \frac{\tilde{N}^{N-1}}{(N-1)!}\,,
\end{align}

where the $N=0$ term vanishes. The sum is the Taylor series of the exponential $e^{\tilde{N}}$, which shows that the noise is indeed additive and uncorrelated when the field is sampled this way. This result has also been tested numerically by sampling ten realizations of the noise and displaying the Fourier analogue of the two terms on the left and right of equation \eqref{eq:shot_noise_proof-2} in figure \ref{fig:shot_noise_proof}. The individual realizations tend to scatter around the theoretical expectation at large scales, while they converge to it when the number of Fourier modes increases, possibly an effect similar to cosmic variance. Plotting the difference between the power spectra of the noisy field and the starting dark matter field, one expects that its value is constant and goes as $1/\bar{n}$ \cite{Feldman1994}. Indeed figure \ref{fig:shot_noise_theory} shows that this is the case. Again, the same behavior as for the previous figure is seen, however at very small scales the noise shows a systematic increase in power compared to theoretical expectation, likely due to aliasing induced in $\delta_m$ when interpolating the N-body dark matter particles onto a regular grid. 

\bibliographystyle{JHEP2015}
\bibliography{main}%{mybib,biblio,references}
\label{lastpage}

\end{document}